\documentclass[12pt,preprint]{aastex}

\slugcomment{}

\shorttitle{\ensuremath{K}-corrections of Type Ia SNe}
\shortauthors{Hsiao et al.}

\begin{document}

\title{\ensuremath{K}-corrections and spectral templates\\of Type Ia supernovae}

\author{
 E.~Y.~Hsiao\altaffilmark{1}, 
 A.~Conley\altaffilmark{2}, 
 D.~A.~Howell\altaffilmark{2}, 
 M.~Sullivan\altaffilmark{2},\\ 
 C.~J.~Pritchet\altaffilmark{1}, 
 R.~G.~Carlberg\altaffilmark{2}, 
 P.~E.~Nugent\altaffilmark{3}, 
 M.~M.~Phillips\altaffilmark{4}
 }

\altaffiltext{1}{Department of Physics and Astronomy, University of Victoria, PO Box 3055, Stn CSC, Victoria, BC V8W 3P6, Canada}
\altaffiltext{2}{Department of Astronomy and Astrophysics, University of Toronto, 50 St. George Street, Toronto, ON M5S 3H4, Canada}
\altaffiltext{3}{Computational Research Division, Lawrence Berkeley National Laboratory, MS 50F-1650, 1 Cyclotron Rd, Berkeley, CA 94720, USA}
\altaffiltext{4}{Las Campanas Observatory, Carnegie Observatories, Casilla 601, La Serena, Chile}

\newcommand{\kcorr}{\ensuremath{K}-correction}
\newcommand{\kcorrs}{\ensuremath{K}-corrections}
\newcommand{\kcorrect}{\ensuremath{K}-correct}
\newcommand{\kcorrected}{\ensuremath{K}-corrected}
\newcommand{\gp}{\ensuremath{g_M}}
\newcommand{\rp}{\ensuremath{r_M}}
\newcommand{\ip}{\ensuremath{i_M}}
\newcommand{\zp}{\ensuremath{z_M}}

\begin{abstract}

With the advent of large dedicated Type Ia supernova (SN\,Ia) surveys, \kcorrs\ of SNe\,Ia and their uncertainties have become especially important in the determination of cosmological parameters. While \kcorrs\ are largely driven by SN\,Ia broad-band colors, it is shown here that the diversity in spectral features of SNe\,Ia can also be important. For an individual observation, the statistical errors from the inhomogeneity in spectral features range from 0.01 (where the observed and rest-frame filters are aligned) to 0.04 (where the observed and rest-frame filters are misaligned). To minimize the systematic errors caused by an assumed SN\,Ia spectral energy distribution (SED), we outline a prescription for deriving a mean spectral template time series which incorporates a large and heterogeneous sample of observed spectra. We then remove the effects of broad-band colors and measure the remaining uncertainties in the \kcorrs\ associated with the diversity in spectral features. Finally, we present a template spectroscopic sequence near maximum light for further improvement on the \kcorr\ estimate. A library of $\sim600$ observed spectra of $\sim100$ SNe\,Ia from heterogeneous sources is used for the analysis.

\end{abstract}

\keywords{cosmology: observations, stars: supernovae}


\section{Introduction}
\label{sec:introduction}

Pioneering SN\,Ia surveys led to the surprising discovery that the cosmic expansion is now accelerating; this acceleration is believed to be driven by dark energy, which constitutes up to $70\%$ of the matter-energy content of the universe \citep{1997ApJ...483..565P,1998Natur.391...51P,1999ApJ...517..565P,1998ApJ...493L..53G,1998AJ....116.1009R,1998ApJ...507...46S}. Although much theoretical work has been done on dark energy, its true nature remains a mystery \citep{2003RvMP...75..559P}. 

The measurement of the equation-of-state parameter can exclude certain theoretical models of dark energy \citep{2001PhRvD..64l3527H}. Large surveys such as the SNLS\footnote{SuperNova Legacy Survey; see http://www.cfht.hawaii.edu/SNLS/} \citep{2006A&A...447...31A} and ESSENCE\footnote{Equation of State: SupErNovae trace Cosmic Expansion; see http://www.ctio.noao.edu/essence/} \citep{2005AJ....130.2453K} aim to constrain the equation-of-state parameter of dark energy with photometric and spectroscopic observations of hundreds of high-redshift SNe\,Ia. With such a large sample size, the systematic uncertainties in \kcorrs\ and photometric calibration become comparable to the intrinsic dispersion in the error budget. Improvements in the \kcorr\ calculations and more sophisticated estimates of the associated errors are thus especially important.

A \kcorr\ converts an observed magnitude to that which would be observed in the rest frame in another filter, allowing the comparison of the brightness of SNe\,Ia at various redshifts \citep{1968ApJ...154...21O,2002astro.ph.10394H}. The \kcorr\ calculations require the SED of the SN\,Ia. A spectral template time series of SN\,Ia is usually used as an assumed SED. This is justified as there exists remarkable homogeneity in the observed optical spectra of ``normal'' SNe\,Ia \citep{1993AJ....106.2383B}. The dependence of \kcorrs\ on the spectral template is large at redshifts where the observed and the rest-frame filters are misaligned. While the observed spectra of ``normal'' SNe\,Ia are apparently uniform, subtle differences in feature strengths and velocities do exist \citep{1995ApJ...455L.147N,2005ApJ...623.1011B,2006MNRAS.370..299H}. How well the spectral template represents the SED of the SNe\,Ia at the misaligned redshifts therefore becomes especially important.

Previous papers on \kcorrs\ of SNe\,Ia developed methods which made it possible to use SNe\,Ia for the determination of cosmological parameters. \citet{1993PASP..105..787H} explored the possibility of using SNe\,Ia as distance indicators and presented single-filter \kcorrs\ for supernovae out to redshift $z=0.5$ using spectroscopic observations of SN 1990N, SN 1991T and SN 1992A. At high redshifts, considerable extrapolation is required for single-filter \kcorrs. \citet{1996PASP..108..190K} developed the cross-filter \kcorr\ to take advantage of the overlap at high redshifts of a rest-frame filter band and a redder observed filter band. \citet[ hereafter N02]{2002PASP..114..803N} presented a SN\,Ia spectral template time series and a recipe for determining \kcorrs. The importance of broad-band colors was particularly emphasized.

While \kcorrs\ are largely driven by SN\,Ia colors, it is shown here that the diversity in spectral features of SNe\,Ia can also be important. This paper investigates the effects of SN\,Ia inhomogeneity on the determination of \kcorrs, paying close attention to the effects of inhomogeneity in spectral features. There are two goals for this paper. The first goal is to demonstrate a method for combining a large and heterogeneous set of observed spectra to make a representative mean spectral template. The second goal is to measure the remaining uncertainties in the \kcorrs\ once the effects of colors have been removed. We test the efficacy of the methods using a large library of $\sim600$ observed spectra of $\sim100$ SNe\,Ia.

In Section~\ref{sec:kcorr}, we outline the method for calculating the \kcorr\ of a SN\,Ia, taking into account the colors of the SN\,Ia. The library of observed SN\,Ia spectra is described in Section~\ref{sec:library}. In Section~\ref{sec:color}, we outline the method for color-correcting a supernova spectrum, a procedure necessary for \kcorr\ calculations and for quantifying the effects of SN\,Ia spectral diversity independent of broad-band colors. Section~\ref{sec:template} provides the prescription for building a mean spectral template time series which has statistically representative spectral features. Section~\ref{sec:error} characterizes the associated \kcorr\ errors; Section~\ref{sec:cosmology} quantifies the effects of these errors on cosmology. In Section~\ref{sec:sequence}, we present a template spectroscopic sequence and its potential for further improvement on the \kcorr\ estimate. 


\section{\kcorrs}
\label{sec:kcorr}

\kcorrs\ of SNe\,Ia rely on the SED of the supernova. High signal-to-noise spectrophotometry of high-redshift SNe\,Ia at multiple epochs is not currently feasible, and it is therefore necessary to use a spectral template time series. 

\begin{figure} 
\plotone{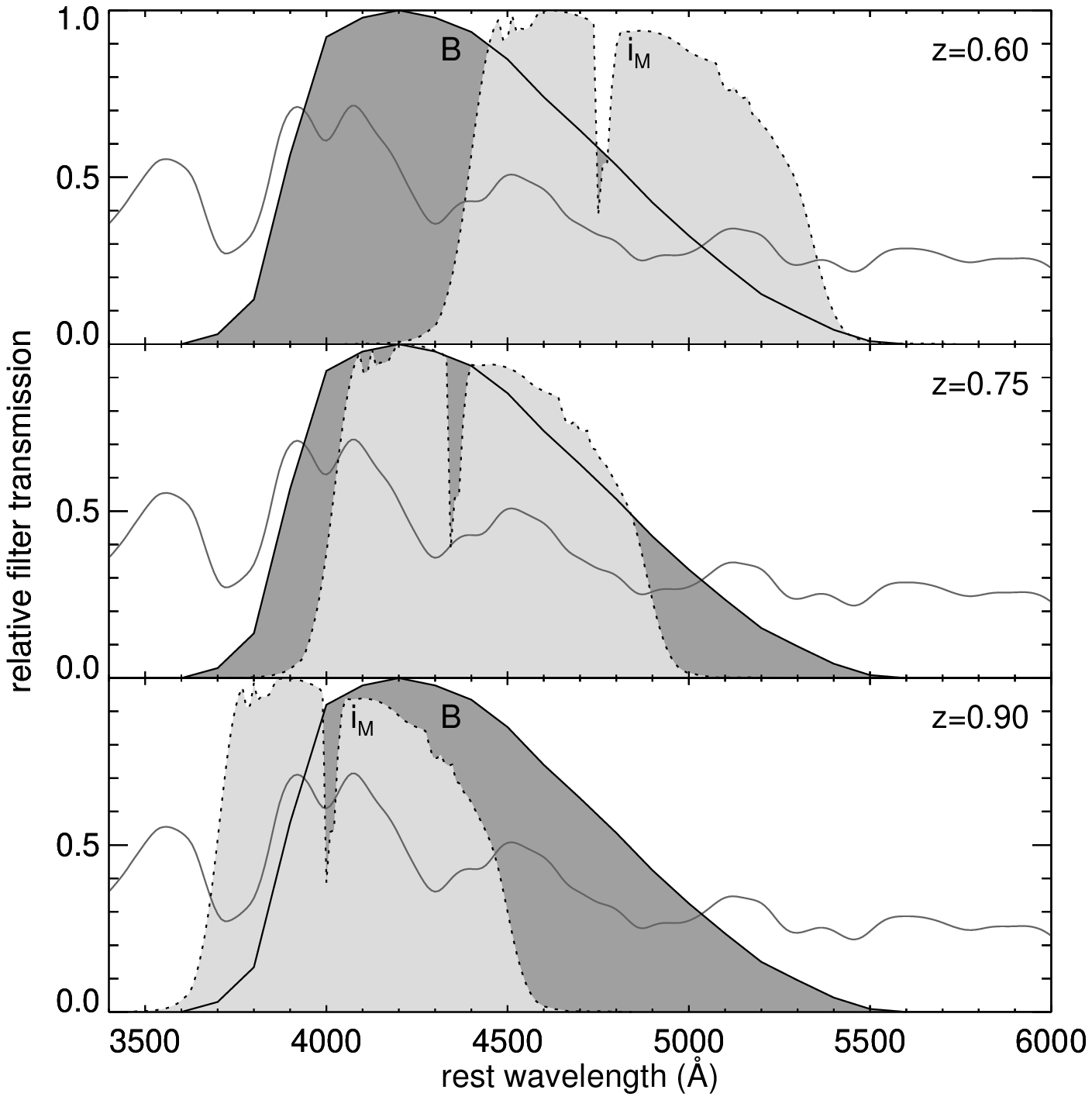}
\caption{The pairing of observed filter \ip\ and rest-frame filter $B$ at redshifts 0.6 (top panel), 0.75 (middle panel) and 0.9 (bottom panel). The solid curves are the transmission of $B$ band, and the dotted curves are the transmission of the deredshifted \ip\ band. The filter bands are misaligned at $z=0.6$ and $z=0.9$. The \kcorrs\ at these redshifts depend heavily on the assumed spectral template.
  \label{fig:align}}
\end{figure}

For a high-redshift supernova, the light of a rest-frame filter band is redshifted to a longer wavelength, and is observed through an observed filter band that overlaps, but is not identical to, the redshifted rest-frame filter band (Figure~\ref{fig:align}). The cross-filter \kcorr\ \citep{1996PASP..108..190K}, $K_{xy}$, allows one to transform the magnitude in the observed filter, $y$, to the magnitude in the rest-frame filter, $x$:

\begin{equation}
\label{eqn:kcorr}
\begin{array}{c}
K_{xy}(t,z,\vec{p})=
-2.5\log \left( \frac
 {\int\lambda T_x(\lambda) Z(\lambda) d\lambda}
 {\int\lambda T_y(\lambda) Z(\lambda) d\lambda} \right) \\	
+2.5\log \left( \frac	
 {\int\lambda T_x(\lambda) S(\lambda,t,\vec{p}) d\lambda}
 {\int\lambda T_y(\lambda) 
 \left[
      S
      \left(
      \lambda/(1+z),t,\vec{p}
      \right)
      /(1+z)
 \right] d\lambda} \right).
\end{array}
\end{equation}

The \kcorr, $K_{xy}$, is a function of the epoch $t$, the redshift $z$, and a vector of parameters, $\vec{p}$, which affect the broad-band colors of the SN\,Ia (e.g., light-curve shape and reddening).  Here, $S$ designates the SED of the supernova, and $\lambda$ designates the wavelength. $T_x$ and $T_y$ denote the effective transmission of the $x$ and $y$ filter bands, respectively. $Z$ denotes the SED for which the $x-y$ color is precisely known. Equation~\ref{eqn:kcorr} essentially compares the supernova fluxes in a rest-frame filter $x$ and a de-redshifted observed filter $y$. Note that the normalization of the SED is arbitrary.

In this paper, we concentrate on the \kcorrs\ to rest-frame $B$ filter band. We use the effective transmission of four MegaCam filters, \gp\rp\ip\zp, as examples of observed filters and the \citet{1990PASP..102.1181B} realization of the Johnson-Morgan $B$ filter. MegaCam is a wide-field imager on the Canada-France-Hawaii Telescope on Mauna Kea. The MegaCam filters are closest to the US Naval Observatory (USNO) filters of \citet{2002AJ....123.2121S} and similar to the Sloan Digital Sky Survey (SDSS) filters. The methods of building a spectral template and estimating errors are general and not restricted to this set of filter bands.

\kcorr\ calculations also depend on the supernova redshift through the alignment of the rest-frame and the observed filter bands. The observed filter band progressively shifts toward bluer parts of the spectrum as a supernova is redshifted to progressively longer wavelengths (from top to bottom panel of Figure~\ref{fig:align}). When the observed and rest-frame filter bands are misaligned, as in the top and the bottom panels of Figure~\ref{fig:align}, the \kcorrs\ are more dependent on the assumed spectral template.

\kcorrs\ largely depend on the broad-band colors of SNe\,Ia and are thus sensitive to anything that affects the continuum of the SED, such as Milky Way and host galaxy extinction, and the intrinsic colors of the supernova. When a spectral template is used as an assumed SED for the supernova, its continuum must be adjusted to have the same colors as the supernova before the \kcorrs\ can be determined. N02 demonstrated that two SNe\,Ia with very different spectral features can yield similar \kcorrs\ when the spectra of the SNe\,Ia are adjusted to the same broad-band colors. Adjusting the colors of the spectral template to match those of the particular SN\,Ia significantly improves the accuracy of the \kcorr\ (N02).

It is an observational fact that SNe\,Ia show a range of feature velocities and strengths \citep[e.g.,][]{2005ApJ...623.1011B}; therefore, it will not generally be true that two SNe\,Ia with identical broad-band colors have identical SEDs. The diversities in individual feature strengths and velocities between SNe\,Ia must affect \kcorr\ calculations at some level. With large, modern surveys, the precision and accuracy requirements for the \kcorrs\ are only increasing. The uncertainties caused by the diversities in spectral features become significant in such cases. This paper will focus on the effects of spectral features, rather than colors, on \kcorrs\ and address the issue by building a statistically representative spectral template.


\section{Library of spectra}
\label{sec:library}

\begin{figure} 
\plotone{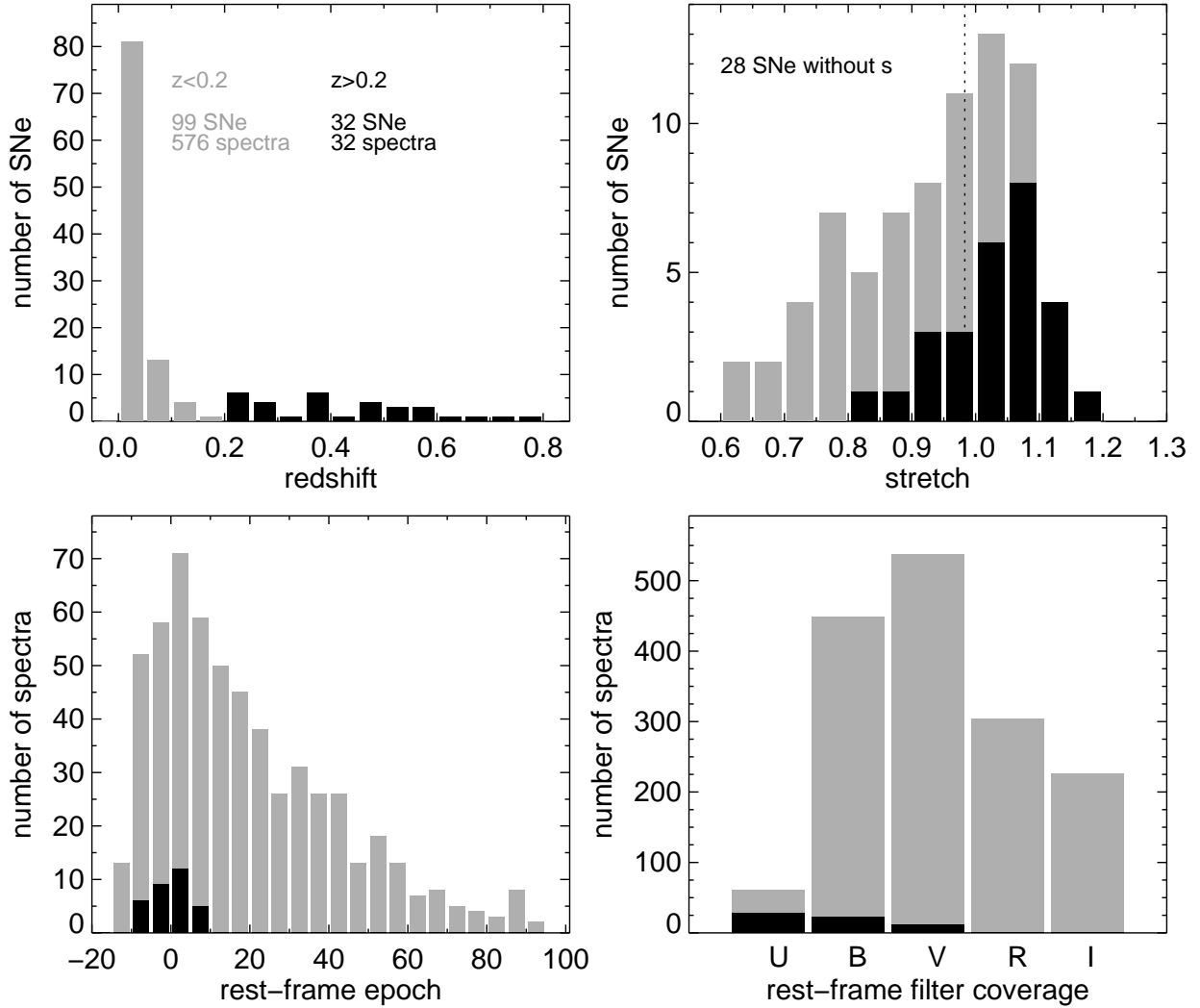}
\caption{Characteristics of our current spectral library. The histograms are separated by redshift at $z=0.2$ to demonstrate the difference in characteristics between the two samples (grey for $z<0.2$ and black for $z>0.2$). The top left panel plots the histogram of the redshifts of the library supernovae. The top right panel plots the histogram of the stretch factors of the library supernovae. Only supernovae with reliable photometry are included in this plot. The median stretch value for the combined sample is marked with a dotted vertical line. The bottom left and right panels plot the histograms of the rest-frame epoch relative to $B$ band maximum light and the rest-frame wavelength coverage of the library spectra, respectively.
  \label{fig:hist}}
\end{figure}

To study the effects of using an assumed time series of spectral templates on the determination of \kcorrs, it is important to use a statistically significant sample of observed SN\,Ia spectra which spans a wide range of supernova properties. We have constructed, and are continuing to update, a library of observed SN\,Ia spectra which includes $\sim600$ spectra from $\sim100$ SNe\,Ia for this study. This paper is primarily a methods paper. Our template spectra, and references to the component spectra that enter into calculating the templates, can be found on the web\footnote{See http://www.astro.uvic.ca/\~{}hsiao/kcorr/}. The spectral template time series is being constantly improved as more and better SN\,Ia spectra become available. The properties of the library as used in this paper are illustrated in Figure~\ref{fig:hist}. 

The observed spectra were gathered from heterogeneous sources with a variety of wavelength coverage, redshifts and quality. A large fraction of the spectra was obtained from public resources, such as the SUSPECT\footnote{See http://bruford.nhn.ou.edu/\~{}suspect/index.html.} database and publications. Other spectra were supplied by past and ongoing supernova surveys, such as the Cal\'an/Tololo Supernova Survey \citep{1996AJ....112.2408H}, the Supernova Cosmology Project \citep{2004AJ....128..387G,2005AJ....130.2278G,2005A&A...430..843L}, the Carnegie Supernova Project \citep{2006PASP..118....2H}, the SNLS \citep{2005ApJ...634.1190H,Bro07} and ESSENCE \citep{2005AJ....129.2352M}. 

Each observed spectrum is inspected by eye for quality. Spectra which have their spectral features dominated by noise are discarded. The spectra taken from ground-based telescopes have the telluric features removed. The edges of each spectrum are inspected and trimmed where the flux calibrations appear unreliable.

The library spectra are overwhelmingly of low redshift SNe\,Ia (top left panel of Figure~\ref{fig:hist}). Spectroscopic observations of low-redshift supernovae are less affected by host galaxy contamination and are generally of higher signal-to-noise than high-redshift ones. On the other hand, high-redshift surveys produce spectra with better observations of the difficult UV region, which is crucial for \kcorrs\ to $B$ at high redshifts. We include spectra from supernovae with redshifts up to 0.8, and work under the assumption that there is no significant evolution of spectral features between low-redshift and high-redshift supernovae \citep{Fol04,2005AJ....130.2788H,2006AJ....131.1648B,Bro07}. The histograms in Figure~\ref{fig:hist} are separated by redshift at $z=0.2$ to demonstrate the difference in characteristics between the two samples.

The library spectra best cover rest-frame $B$ and $V$ filter bands (bottom right panel of Figure~\ref{fig:hist}). We will focus our analysis on \kcorrs\ to the $B$ band, since it is the most important and commonly used band in current cosmological analysis. When the observed and the rest-frame $B$ filter bands are misaligned, the spectral features in the neighboring $U$ and $V$ filter bands are especially important. The relative shortage of UV spectra is slightly improved by the inclusion of high-redshift spectra.

The rest-frame epochs of the spectra are relative to $B$ band maximum light and are obtained mainly from the light curves of the supernovae. When the photometry of a SN\,Ia is unavailable or unreliable, an estimate of the epoch is made by fitting the spectrum to other spectra of known ages. The distribution of the epochs peaks at maximum light (bottom left panel of Figure~\ref{fig:hist}). There are no late time spectra for the high-redshift sample.

The correlation between the peak luminosity and the light-curve shape \citep{1993ApJ...413L.105P} is a useful parameter to characterize the diversity of the library SNe\,Ia. For each SN\,Ia with well sampled light curves, a stretch factor is computed. The stretch factor, $s$, linearly stretches or contracts the time axis of a template light curve around the epoch relative to maximum $B$ band light to best fit the observed light curve \citep{1997ApJ...483..565P,1999ApJ...517..565P,2001ApJ...558..359G}. The stretch factors of the library SNe\,Ia span a wide range of values with most of the supernovae around stretch of unity (top right panel of Figure~\ref{fig:hist}). The high-redshift sample has a slightly higher median stretch ($s=1.049$) than the low redshift sample ($s=0.967$). This is probably caused by the preferential selection of brighter (higher stretch factor) SNe\,Ia at high redshifts. The median stretch factor for the combined sample is 0.983.

Even though there are established relations between light-curve shape and intrinsic SN\,Ia colors \citep[e.g.,][]{1996ApJ...473...88R}, the spectral feature strengths and velocities of SNe\,Ia do not follow a simple relation at all ages and wavelengths. SNe\,Ia with the same light-curve shape do not necessarily exhibit the same spectral feature strengths; therefore, the diversity of spectral features cannot be completely characterized by one parameter alone. We can only strive for a truly diverse and representative library by including as many observed spectra from as many SNe\,Ia as possible.


\section{Color-correction}
\label{sec:color}

\begin{figure} 
\plotone{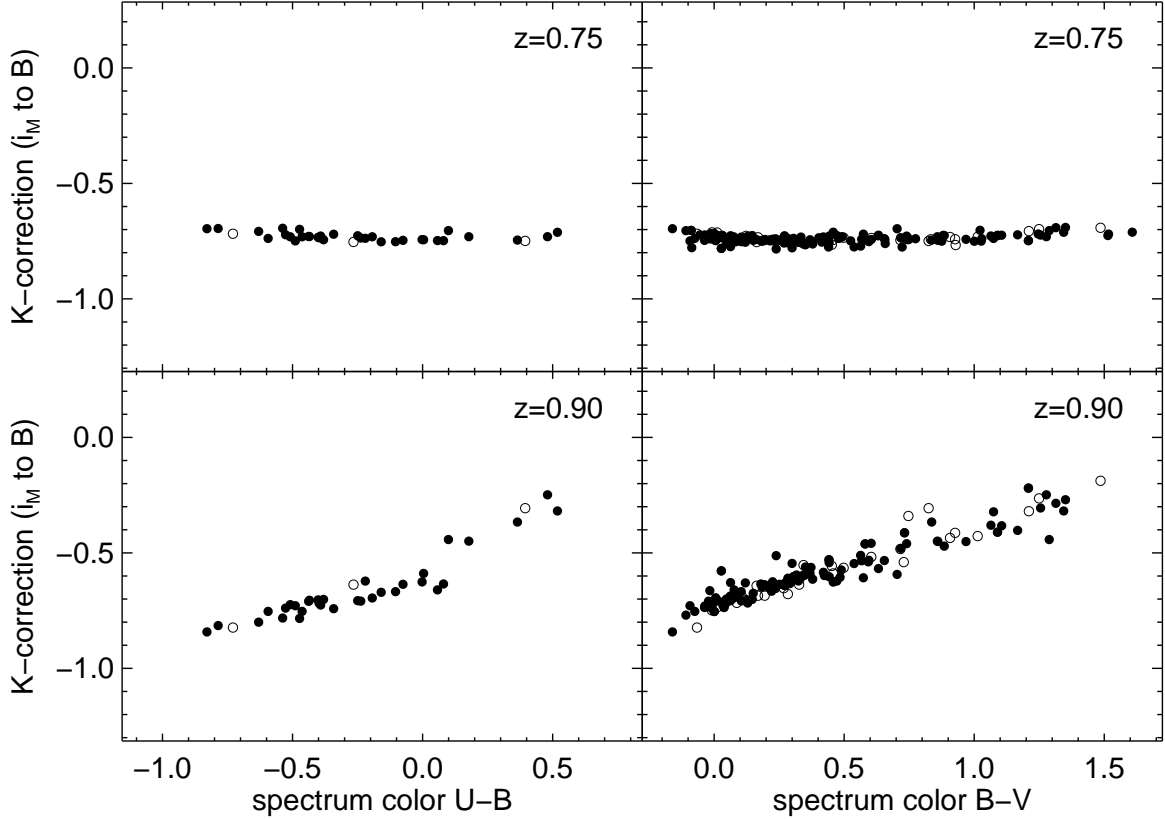}
\caption{\kcorrs\ from observed \ip\ band to rest-frame $B$ band as a function of broad-band colors, $U-B$ (left panels) and $B-V$ (right panels). The \kcorr-color relations are plotted at the redshifts $z=0.75$ (top panels) and $z=0.9$ (bottom panels). The \ip\ and $B$ filter bands are aligned at $z=0.75$ and misaligned at $z=0.9$. All the library spectra with adequate wavelength coverage are included. Each point represents one library spectrum. Filled and open circles represent normal and peculiar SNe\,Ia, respectively. The diversity in colors mostly reflects the time evolution of supernova colors. The scatter reflects the differences in spectral features.
  \label{fig:kc}}
\end{figure}

To first order, \kcorrs\ are driven by the diversity of SN\,Ia colors, whether this is attributed to extinction or to the variations of SN\,Ia SED with age and light-curve shape. We illustrate the relation between \kcorrs\ and supernova colors in Figure~\ref{fig:kc}. The \kcorrs\ are calculated for each library spectrum. The spectrum colors are determined from synthetic photometry on the actual observed spectrum.

We use the colors from the spectra themselves instead of photometric colors for the following reasons. Many library spectra are not spectrophotometric; therefore, using colors from photometry introduces errors which do not originate from supernova feature inhomogeneity. This choice was made also for practical reasons as the photometry of the library supernovae is often unavailable or unreliable. 

In Figure~\ref{fig:kc}, we demonstrate the relation between the \kcorr\ and broad-band colors at two redshifts for $\ip \mapsto B$. At a redshift of $0.75$ where the $B$ and \ip\ filter bands are aligned (middle panel of Figure~\ref{fig:align}), the \kcorrs\ are mostly constant with respect to the colors of the supernovae (top panels of Figure~\ref{fig:kc}). At $z=0.9$, the filter bands are misaligned (bottom panel of Figure~\ref{fig:align}), and the \kcorrs\ show a strong dependence on broad-band colors with a larger scatter around the trend (bottom panels of Figure~\ref{fig:kc}). At these misaligned redshifts, the \kcorr\ involves more extrapolation and hence is more reliant on the details of the SED. The colors of the SED describe the bulk of the relation. The remaining scatter can be attributed to the inhomogeneity in spectral features, which is the focus of this paper.

How should one allow for color variations when calculating \kcorrs? N02 utilized the reddening law of \citet{1989ApJ...345..245C} to perform the color-correction on the spectral template. This approach was also adopted by the High-z Supernova Search Team \citep{1998AJ....116.1009R,2003ApJ...594....1T}. This choice is based on the observation that SN\,Ia color relations roughly follow the reddening law. The correction applies a slope as a function of wavelength to the spectral template in order to obtain the desired $B-V$ color.

Large surveys, such as the SNLS, produce well sampled multi-filter light curves and in turn produce multi-color information for most SNe\,Ia. To utilize fully the multi-color information, we have developed a ``mangling'' (color-correction) function in lieu of the reddening law slope correction. The mangling function is similar to the treatment of spectral template color adjustment in \citet{2004ApJ...607..665R}. It defines splines as a function of wavelength, with knots located at the effective wavelengths of the filters. Non-linear least-squares fitting\footnote{See http://cow.physics.wisc.edu/\~{}craigm/idl/idl.html.} \citep{Mor80} is used to determine the spline which smoothly scales the spectral template to the correct colors.

\begin{figure} 
\plotone{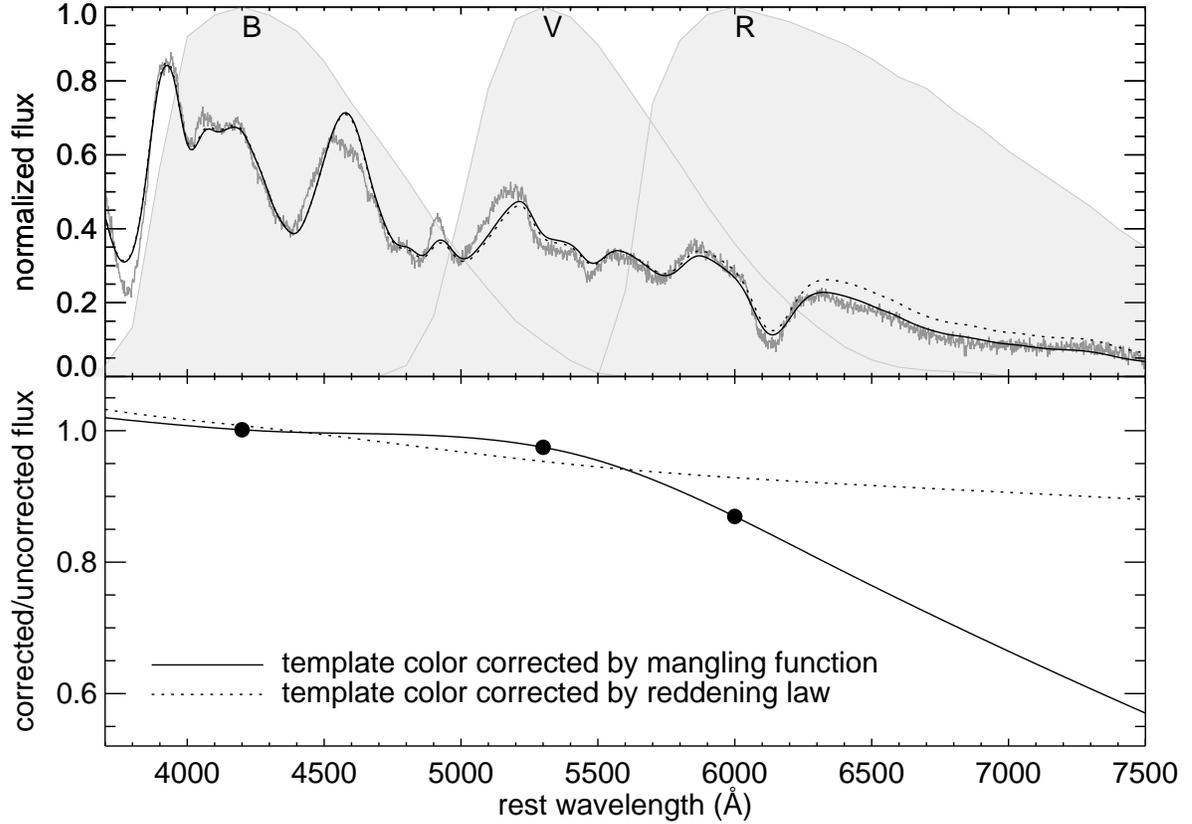}
\caption{An example of color-correction which contrasts the two color-correction methods: reddening law slope correction and mangling function. The spectral template is color-corrected to match the colors of an observed spectrum. The top panel shows three spectra: the observed spectrum (grey), the spectral template corrected by the mangling function (black solid) and the spectral template corrected by the reddening law (black dotted). The bottom panel shows the color-correction scale for the mangling function (black solid) and for the reddening law (black dotted). The black dots locate the spline knots for the mangling function. The mangling function adjusts the continuum using all the color information supplied, $B-V$ and $V-R$.
  \label{fig:mangeg}}
\end{figure}

Figure~\ref{fig:mangeg} shows an example of color-correction which contrasts the mangling function and the slope correction of the reddening law. The two methods exhibit similar behavior between the $B$ and $V$ filter bands. When the $V-R$ color information is included, the scale in the $R$ band determined by the mangling function diverges from a slope correction to incorporate the extra color information. The mangling function anticipates the possibility that the reddening law does not completely characterize all the colors of a SN\,Ia; it effectively pieces together segments of slope corrections smoothly to satisfy the multi-color information. 

The variations in large features of SNe\,Ia, such as the Ca H\&K feature in the UV, can significantly alter the colors of the supernovae. It is difficult to disentangle color and feature variations in these cases; however, the results of the analysis in this paper are unaffected as long as the distinctions between the effects of color and feature are consistently defined. Before spectra are compared with one another, they are color-corrected to the same broad-band colors using the mangling function described here. This procedure consistently defines an effective ``continuum'' for all spectra and enables the study of the effect of spectral features on \kcorrs\ independent of colors, whether the color differences are caused by extinction, intrinsic colors or observational effects. 

Returning to Figure~\ref{fig:kc}, we can see that it illustrates not only the general trend of \kcorr\ in terms of supernova colors, but also the scatter around the trend which is caused by feature variation between supernovae. The scatter becomes more significant as the rest-frame filter band is redshifted away from the observed filter band (from top to bottom panels of Figure~\ref{fig:kc}) showing the heavier reliance on the spectra and the inhomogeneity of the spectral features. The goal now is to construct a spectral template time series that is representative of the observed SN\,Ia spectral features and to characterize the errors associated with the spectral inhomogeneity observed in the scatter.


\section{Constructing a spectral template}
\label{sec:template}

The spectra of normal SNe\,Ia are remarkably uniform in the time evolution of their spectral features \citep{1993AJ....106.2383B}, but some differences do exist. Some of these differences are associated with the variation in light-curve shapes \citep{1995ApJ...455L.147N,2005ApJ...623.1011B,2006MNRAS.370..299H}, while the origins of others remain unknown.

N02 introduced a spectral template time series that is based on one well-observed supernova spectrum for each epoch and wavelength interval. The template is constructed by assembling well-observed spectra in a two-dimensional grid of flux as a function of epoch and wavelength. The temporal gaps are filled with a simple interpolation.

Supernova spectra with the same broad-band colors do not necessarily have the same spectral features inside a broad filter band. The spectral features of the template used for \kcorr\ calculations of a particular set of SNe\,Ia should be representative of the population. The template constructed by the method outlined in N02 can cause systematic errors if the template represents the extremes in spectral features. In this section, we outline a prescription for constructing a spectral template time series of SN\,Ia with representative spectral features.

Simply averaging a large number of spectra tends to wash out spectral features, when there exists a sizeable range of feature velocities. Instead, we adjust the features of a base template to the weighted mean of the measured feature strengths from a large sample of observed spectra. The procedures of building the spectral template time series have been built into a fully automated IDL program. The program can incorporate new additions to the library, adjust the grid size accordingly and create a new spectral template. We outline the procedures in the following summary, while the details of each step are described in the following subsections:

\begin{enumerate}
\item Assign a two-dimensional grid of epoch and wavelength bins.
\item Color-correct each library spectrum to the same broad-band colors in each epoch bin.
\item Measure the feature strength in each wavelength bin of each library spectrum.
\item Assign weights to each feature strength of each library spectrum.
\item Determine an effective epoch for each epoch bin.
\item Determine a weighted mean feature strength for each grid element.
\item Build a base template from a subset of library spectra.
\item Adjust the base template spectrum to the weighted mean feature strengths using the mangling function.
\end{enumerate}

\subsection{Epoch bins}
\label{sec:template:epoch}

We first divide the library spectra into epoch bins. When assigning an epoch bin, two competing factors affecting the bin sizes must be considered. First, a statistically significant sample size of spectra in each epoch bin is required to obtain representative characterizations of the spectral features; larger bin sizes yield more spectra per bin. Second, the temporal evolution of the spectral features within each bin should be kept as small as possible; small epoch bin sizes are preferred in this case. The most direct way to improve this situation is to increase the number of library spectra such that the epoch bin sizes can be narrowed without sacrificing the generality of the sample in each bin. For the current sample size, we cannot afford to set a bin size of one day for all epochs. Instead, we adopt variable bin sizes.

\begin{figure} 
\plotone{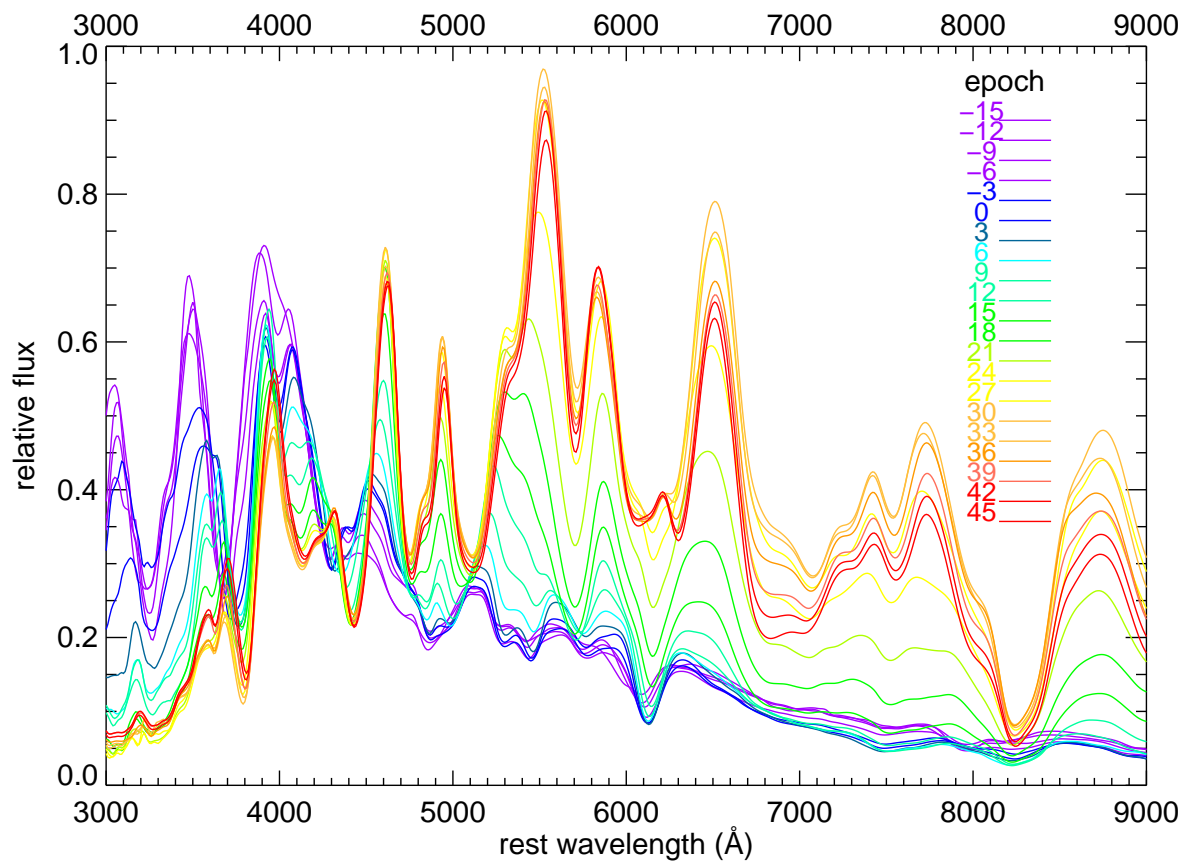}
\caption{An illustration of the time evolution of spectral features of SNe\,Ia. A spectral template is plotted from $t=-15$ to $t=48$ relative to maximum B band light. Template spectra at different epochs are normalized to the same B band flux to emphasize the evolution of the spectral features. Spectral features of SNe\,Ia evolve rapidly around maximum light and slow down past $t=30$. This emphasizes the importance of small epoch bins near maximum light.
  \label{fig:lineid}}
\end{figure}

The temporal evolution of spectral features is plotted in Figure~\ref{fig:lineid}. The SEDs at different ages are normalized to the same $B$ band flux to emphasize the evolution of the spectral features. The spectrum of a SN\,Ia evolves rapidly from the time of explosion to around 20 days past maximum light. These epochs are also the most important when fitting light curves. The epoch bin sizes for these epochs are kept small such that the effects of temporal evolution on spectral features are minimized. The temporal evolution of feature shapes in the $B$ and $V$ bands slows down considerably beyond the age of 30 days. At these epochs, we adopt larger bin sizes to compensate for the small number of spectra without losing too much information about temporal evolution. Table~\ref{tbl:epoch} lists the epoch bins adopted for this particular set of library spectra.

It is unclear whether the SED of a SN\,Ia with stretch $s$ at time $t$ relative to maximum light corresponds to the template of stretch $s=1$ at age $t$ or $s \times t$. This uncertainty however has little effect on our template building, as the large epoch bins at late times render the two options indistinguishable.

\subsection{Measurements of feature strengths}
\label{sec:template:measure}

\begin{figure} 
\plotone{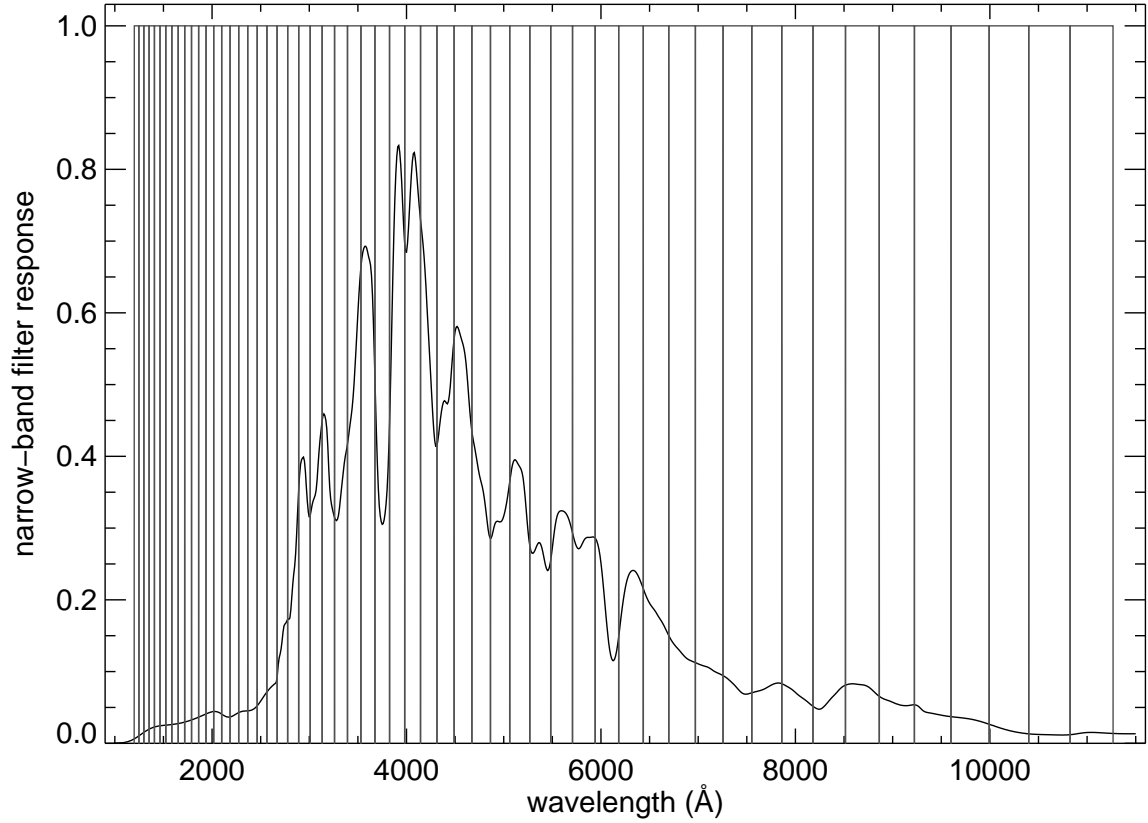}
\caption{The defined narrow-band filters for the measurement of spectral feature strengths. The filters are defined to have logarithmic bandwidths with a ratio of $\Delta \lambda/\lambda=0.04$. The bandwidths are on the order of the sizes of SN\,Ia spectral features.
  \label{fig:nfilters}}
\end{figure}

We define a set of artificial non-overlapping narrow-band pass filters for measuring the spectral feature strengths (Figure~\ref{fig:nfilters}). The bandwidths of the narrow-band filters are logarithmic in wavelength and are designed to be on the order of the sizes of supernova features. We adopt a ratio of $\Delta \lambda/\lambda=0.04$, which yields a bandwidth of 160\AA\ at $\lambda=4000$\AA\ and a bandwidth of 280\AA\ at $\lambda=7000$\AA. The large expansion velocities of SNe\,Ia mean that the bandwidth of the narrow-band filters can be set large enough such that modest noise in observed spectra has minimal effect on the measurements of feature strengths.

Before the feature strengths are measured, each library spectrum is color-corrected using the mangling function to have the same broad-band colors as a supernova with a set stretch value (conventionally $s=1$). The templates of \citet{2003ApJ...598..102K} are used to determine the colors. It is worth noting here that the procedure for calculating \kcorrs\ presented in this paper is independent of the stretch-color relation. Even though the relation at a particular stretch value is used here as the standard colors for the library spectra, the spectral template would later be color-corrected to match the colors of the particular SN\,Ia in question. The color-correction procedure provides a consistent ``continuum'' for all the library spectra and removes the dependence on color varying factors, such as stretch, extinction and flux calibration. The spectral diversity of the library SNe\,Ia can now be adequately quantified independent of broad-band colors.

\begin{figure} 
\plotone{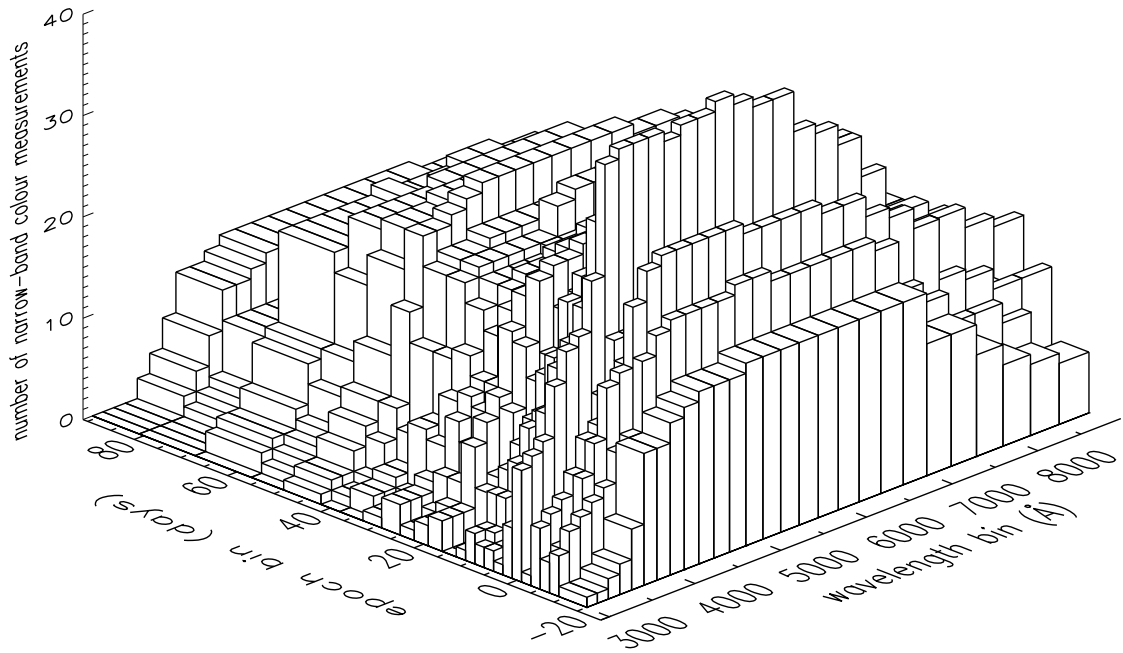}
\caption{A schematic of the two-dimensional grid defined by the epoch bins and the narrow-band filters. The number of narrow-band color measurements in each grid element is plotted in the third dimension.
  \label{fig:grid}}
\end{figure}

The relative flux between neighboring features of a library spectrum is measured as the narrow-band colors from synthetic photometry using neighboring narrow-band filters. Measuring relative feature strengths eliminates the need for the absolute flux of the spectra. The narrow-band colors are organized in a two-dimensional grid of wavelength and epoch. The sizes of the grid elements are defined by narrow-band filter bandwidths and epoch bins. Figure~\ref{fig:grid} illustrates a schematic of the grid and specifies the number of narrow-band color measurements for each grid element. A typical grid element in the $B$ and $V$ filter bands has $\sim20$ measurements.

\subsection{Weighting feature strength measurements}
\label{sec:template:weight}

The heterogeneous nature of our library means that we need to consider the differences between the spectra carefully. Before the narrow-band colors are averaged in each grid element, some simple weighting schemes are applied to each spectrum to ensure that the resulting template is not dominated by peculiar spectra, supernovae with many observations or supernovae with extreme stretch values. Weights as a function of wavelength are also assigned to deal with the differences in the spectral coverage and the location of the telluric lines.

We chose to include spectroscopically peculiar SNe\,Ia for the following reasons. Including peculiar spectra in the template and the analysis gives a more complete description of the population of the SNe\,Ia. Moreover, people tend to make more detailed observations of more peculiar SNe\,Ia. Including these spectra helps to increase the number of spectra in wavelength and epoch intervals which are rarely covered. To adequately include these peculiar spectra in the spectral template, we adopt the weight, $w_{type}$. For the peculiar spectra, we chose a relative weight of one fifth that reasonably characterizes the intrinsic fraction of peculiar supernovae in the SN\,Ia population \citep{2001PASP..113..169B,2001ApJ...546..734L}. For a grid element with $N_{normal}$ normal spectra and $N_{peculiar}$ peculiar spectra, the weight for the type of spectra is:

\[
w_{type} = \left\{ \begin{array}{ll}
                   1/N_{normal}     & \mbox{for normal SNe\,Ia} \\
		   1/5 N_{peculiar} & \mbox{for peculiar SNe\,Ia} 
		   \end{array}
           \right. 	     
\]

For a SN\,Ia with multiple narrow-band color measurements in a grid element, the measurements are averaged and weighted as one measurement ($w_{multiplicity}$).

Narrow-band color measurements at the edges of a library spectrum are assigned less weight than the measurements at the center. This step goes beyond the edge trimming, described in Section~\ref{sec:library}, in minimizing the effects of miscalibration at the edges of observed spectra. It also gives higher weights to library spectra which have wider wavelength coverage and thus have their continua better adjusted by the mangling function. The weight for the filter coverage, $w_{coverage}$, is a triangular or trapezoidal function in wavelength space with a value of unity at the effective wavelengths of the available filter coverage bands and one fifth at the ends of the spectrum.

Even though the telluric features for each spectrum taken at ground-base telescopes are removed, we apply lower weights at the location of the telluric features to account for the errors associated with this procedure. The diversity of redshifts in our library means that most of the grid elements would have at least a few feature strength measurements uncontaminated by telluric features. The weight assigned for telluric line contamination, $w_{telluric}$, is inversely proportional to the equivalent width of a telluric feature model inside the two adjacent narrow-band filters in question.

To ensure that a grid element is not dominated by supernovae of extreme stretch values, the narrow-band color measurements are weighted such that the weighted mean of the stretch values in each grid element is close to a representative stretch value $s_{effective}$ (conventionally unity). We use a normal distribution centered on the effective stretch, $s_{effective}$, with a sigma, $\sigma_{s}=0.2$:

\[ 
w_{stretch} = \exp\left[{\frac{(s_{SN}-s_{effective})^2}{2\sigma_{s}}}\right]
\]

For SNe\,Ia without stretch factors, the mean weight for stretch is adopted for each grid element. Note that we can build a template which has spectral features which are representative for any effective stretch value by choosing $s_{effective}$. Using spectral template with different effective stretch can be effective if spectral features truly vary with stretch. Previous studies have shown that this is the case only for certain epochs and wavelength regions. We will explore the topic of template spectroscopic sequence further in Section~\ref{sec:sequence}.

An effective epoch, $t_{effective}$ is then determined for each epoch bin using the epochs of library spectra, $t_i$, and their weights, $w_i^{\prime}$:

\[
t_{effective} = \frac{\sum_i w_i^{\prime}t_i}{\sum_i w_i^{\prime}}
\]

The letter $i$ denotes the $i$th spectrum in the epoch bin in question. The weight, $w_i^{\prime}$ is the product of the weights for supernova type, multiplicity, wavelength coverage, telluric features and stretch. The resulting effective epochs for our current library are listed in Table~\ref{tbl:epoch}. Within an epoch bin, library spectra with epochs $t_{SN}$ progressively closer to the effective epoch are then assigned progressively higher weights. We use a normal distribution centered on the effective epoch, $t_{effective}$, with a sigma, $\sigma_{t}$, corresponding to the size of the epoch bin:

\[
w_{epoch} = \exp\left[{\frac{(t_{SN}-t_{effective})^2}{2\sigma_{t}}}\right]
\]

The total weight for the $i$th library spectrum can then written as:

\[
w_i = w_{type,i} \times w_{multiplicity,i} \times w_{coverage,i} \times w_{telluric,i} \times w_{stretch,i} \times w_{epoch,i}
\]

Assuming that the spectral features of SNe\,Ia depend on the stretch parameter at some level, weighting feature strength measurements in terms of epoch and stretch helps us to disentangle the two effects on spectral features in an epoch bin of finite size.

\subsection{Building the spectral template}
\label{sec:template:build}

The weighted mean of the narrow-band color measurements, $c$, is determined for each grid element (in epoch and in wavelength) using the total weight, $w_i$, and the narrow-band color measurements $c_i$ for the $i$ spectrum: 

\[
c = \frac{\sum_i w_ic_i}{\sum_i w_i}
\]

The grid of weighted mean narrow-band colors in epoch and wavelength now characterizes the spectral features of the desired SN\,Ia spectral template time series. 

It is important to start with a base template spectrum with the correct feature shapes. We build a base template spectrum at the effective epochs using a selected set of high signal-to-noise spectra from the library. The coverage in epoch and wavelength is kept below three spectra per day per wavelength to avoid washing out the spectral features. Each library spectrum is corrected to the same colors and the same $B$ magnitude at each epoch. Gaussian weights centered on the effective epoch are applied to the flux. The weighted mean flux is then computed for each effective epoch.

The grid of weighted mean narrow-band colors, $c$, as a function of wavelength and epoch is now used as the input color information for the mangling function described in Section~\ref{sec:color}. Each spectral feature of the base spectrum is adjusted by the mangling function to have the weighted mean strength of the library spectra. The adjusted template at the effective epochs is then smoothly interpolated to an integer epoch grid from $t=-19$ to $t=85$ at each wavelength. Finally, the spectral template is color adjusted to a light-curve template.

Note that adjusting the flux and the colors of the spectral template to a particular set of template light curves is optional for the following reasons. The constant of normalization for the template SED is cancelled in Equation~\ref{eqn:kcorr}. Furthermore, the broad-band colors of the spectral template is later adjusted to match the colors of the SN\,Ia in question. Here, we choose to adjust our spectral template to a light-curve template to allow for a more natural interpolation of flux between epochs.


\section{Quantifying \kcorr\ errors}
\label{sec:error}

In this section, we characterize the errors associated with the use of an assumed SN\,Ia spectral template time series when calculating the \kcorrs\ of SNe\,Ia. \kcorr\ errors are quantified here as the differences between the \kcorr\ calculated using an actual supernova spectrum $S^{S}$ and that calculated using an assumed template spectrum $S^{T}$, where the continuum of $S^{T}$ is adjusted with the mangling function to have the same broad-band colors as $S^{S}$ (Section~\ref{sec:color}). From Equation~\ref{eqn:kcorr}, we obtain:

\begin{equation}
\label{eqn:error}
\begin{array}{c}
\Delta K_{xy} \equiv
    2.5 \log
    \left( 
    \frac
        {\int \lambda S^{S}(\lambda) T_x(\lambda) d\lambda}
        {\int \lambda S^{T}(\lambda) T_x(\lambda) d\lambda}
    \right)\\
    - 2.5 \log
    \left( 
    \frac
        {\int \lambda S^{S}(\lambda/(1+z)) T_y(\lambda) d\lambda}
        {\int \lambda S^{T}(\lambda/(1+z)) T_y(\lambda) d\lambda}
    \right).
\end{array}
\end{equation}

This definition of \kcorr\ errors removes the dependence on the zero points and focuses on the effects of an assumed spectral template and the alignment of the rest-frame and observed filters. The first term in Equation~\ref{eqn:error} provides the normalization such that the two spectra have the same flux in the rest-frame filter band. The second term then yields the magnitude difference caused by the spectral feature differences between the supernova spectrum and the template. 

Here, we utilize the library spectra again to quantify the \kcorr\ errors. The library spectra are artificially redshifted from $z=0$ to $z=1.2$ to determine the \kcorr\ as a function of redshift. The \kcorr\ differences between the template and each library spectrum are then determined for all redshifts with various combinations of rest-frame and observed filter bands. We then define the statistical error as the \kcorr\ differences added in quadrature at each redshift and the systematic error as the average of the \kcorr\ differences at each redshift.

When a relation between light-curve shape and color \citep[e.g.,][]{1996ApJ...473...88R} is used, one has available the predictions of multiple rest-frame color information from the fit to the observed light curve. The spectral template can be color-corrected using the predicted rest-frame colors in this case. On the other hand, when multi-color observations of a SN\,Ia are available, one may choose to use the more empirical approach of correcting the spectral template directly to the observed colors. We explore the \kcorr\ errors from both color-correction options in the following subsections.

\subsection{Correcting the template using model rest-frame colors}
\label{sec:error:rest}

We first look at the case where the colors of the spectral template are corrected by the rest-frame colors. When determining the \kcorr\ to a rest-frame filter, the neighboring spectral regions become important when the rest-frame and observed filter bands are misaligned. To determine the errors caused by the inhomogeneity in spectral features, it is important to use spectra which have adequate wavelength coverage and can be accurately color-corrected by the mangling function. We took all the library spectra which have $U$, $B$ and $V$ filter band coverage (51 spectra, about 8\% of our current library) to characterize the errors in \kcorr\ for the rest-frame $B$ filter band. The selected library spectra range in epoch from $t=-15$ to $t=65$. The rest-frame colors of the library spectra are measured by synthetic photometry. Before the \kcorr\ difference between a library spectrum and the template is calculated, the template spectrum is color-corrected to have the same $U-B$ and $B-V$ colors as the library spectrum.

We randomly select half of the library spectra with $U$, $B$ and $V$ band coverage to calculate the \kcorrs. We use the other half and the rest of the spectra without complete $U$, $B$ and $V$ coverage to make the spectral template, such that the spectral template is always independent of the library spectra used to examine the \kcorr\ differences. The process of random selection, building a template with one half of the sample and calculating the \kcorr\ errors with the other half was repeated $100$ times.

\begin{figure}
\plotone{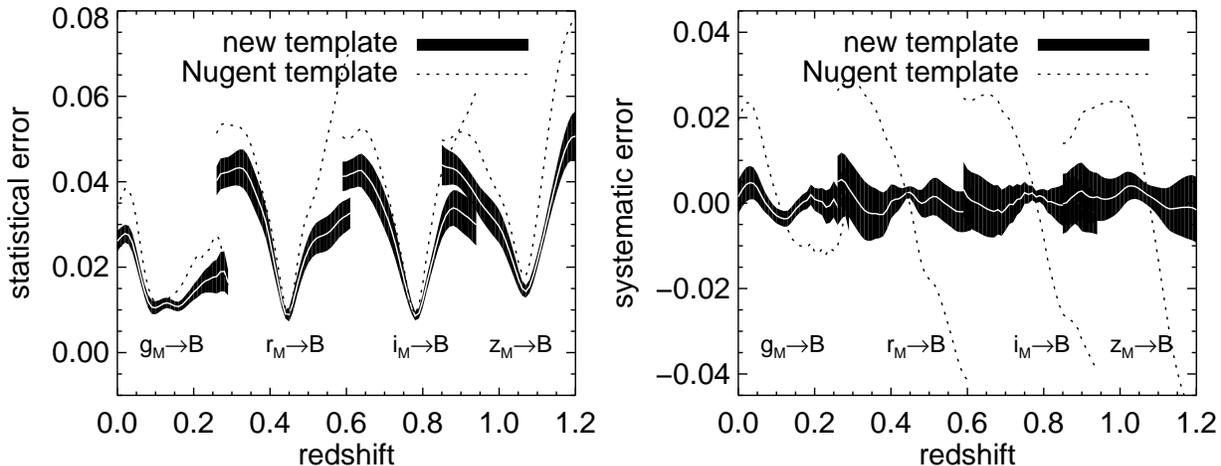}
\caption{\kcorr\ errors to the $B$ band where the template spectrum is color-corrected with rest-frame colors. The errors of the new template and the revised template of N02 are represented with solid black and dotted curves, respectively. The statistical errors are plotted in the left panel, and the systematic errors in the right panel. The thickness of the black solid curve shows the dispersion from the random selection process and the thin white curve is the mean. The errors presented here are for individual observations on a light curve.
  \label{fig:krest}}
\end{figure}

We plot the statistical and systematic errors for \kcorrs\ from observed \gp\rp\ip\zp\ to rest-frame $B$ in Figure~\ref{fig:krest}. The solid black curve is from the new template building method presented here; the dotted curve is from the revised template of N02. The thickness of the black curve shows the dispersion from the random selection process and the thin white curve is the mean. We assume here that the revised template of N02 is independent of the 51 spectra used for the error estimate.

Note that the errors presented here are for individual observations. Combining multiple data points to derive overall light-curve parameters should reduce these errors (Section~\ref{sec:cosmology}). The statistical errors are a function of redshift. They follow the pattern of a minimum at the redshift where the observed and the rest-frame filter bands are aligned and two maxima where the two filter bands are misaligned. The pattern repeats as a function of redshift and as one observed filter is switched to another. At a redshift where the filter bands are misaligned, the \kcorr\ error for a single observation on a light curve is important at the $0.04$ mag level. Because the spectra are compared with consistent broad-band colors, we argue that the $0.04$ mag error is mostly due to the inhomogeneity in spectral features of SNe\,Ia.

The right panel of Figure~\ref{fig:krest} shows that the new template causes lower systematic errors than the revised template of N02. This is because the new template has more representative spectral features for the 51 spectra considered here. This implies that the N02 template is only representative for the handful of spectra used to make it. The method presented in Section~\ref{sec:template} allows for the inclusion of a large sample of spectra for building a spectral template. Assuming that our current library is a more representative sample than the sample used by N02, then the new template will reduce the systematic errors caused by spectral features.

Note that we do not expect the mean systematic error to be identically zero in these plots. An ideal spectral template for this particular population of spectra would yield distributions centered on zero at all redshifts and epochs. The deviation from zero for the new template reflects the fact that only 51 out of the $\sim600$ library spectra were used (due to the wavelength coverage constraint) to estimate the errors and the requirement that the sample used to build the template is disjoint from that used to perform the test. They are consistent with zero within the errors.

When a SN\,Ia lies at a redshift where the observed and rest-frame filter bands are misaligned, all the observations on the light curve would have larger \kcorr\ errors. The larger errors are attributed to the larger extrapolation and the heavier reliance on the spectral template. Multiple epochs may reduce the errors, but the redshift dependent nature remains. SNe\,Ia at different redshifts would have different \kcorr\ errors. This redshift-dependent feature of the \kcorr\ error must be considered in the determination of cosmological parameters.

\subsection{Correcting the template using observed colors}
\label{sec:error:observed}

Here we perform the same error analysis as Section~\ref{sec:error:rest}, but the template spectra are color-corrected using ``observed'' colors. The observed colors are obtained from the synthetic photometry using deredshifted \gp\rp\ip\ observed filter bands. As the redshift of a supernova spectrum increases, the observed filter bands cover progressively bluer regions of the spectrum. The locations of the spline knots for the mangling function are not fixed as in the previous subsection, but instead depend on the redshift of the supernova. 

\begin{figure}
\plotone{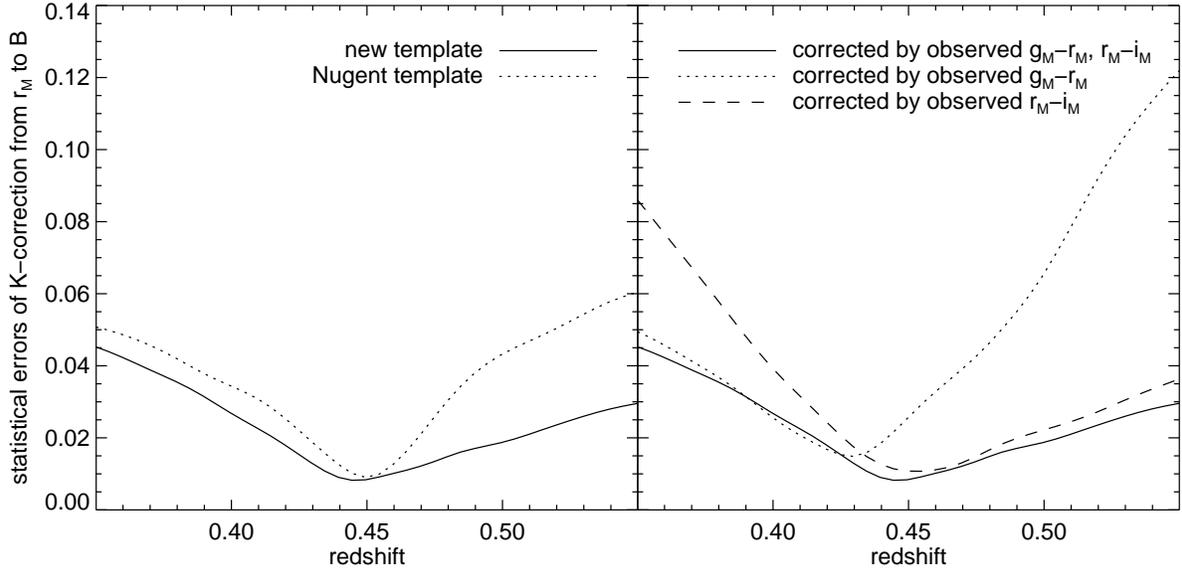}
\caption{Statistical errors of \kcorr\ from observed \ip\ band to rest-frame $B$ band where the template spectrum is color-corrected with ``observed'' colors. The observed colors are from the synthetic photometry using the observed filter bands \gp\rp\ip. The left panel compares the errors of the revised template of N02 and the new template. Both templates are color-corrected by observed $\gp-\ip$ and $\ip-\rp$ colors before the calculation. The right panel shows the effect of the availability of observed colors on the errors. The solid, dotted and dashed curves represent the errors of the new template corrected by both $\gp-\rp$ and $\rp-\ip$ colors, $\gp-\rp$ color alone and $\rp-\ip$ color alone, respectively.
  \label{fig:kobsv}}
\end{figure}

Because the observed filter bands are shifted as a function of redshift, spectra with wide wavelength coverage are required for the analysis. There are 12 spectra with adequate wavelength coverage from seven SNe\,Ia, SNe 1981B, 1989B, 1990N, 1990T, 1991T, 1991bg, 1992A, covering ages from $-14$ to $37$ days relative to maximum light. The spectral template built for this analysis excludes these 12 spectra, such that the spectral template is independent of the these spectra used for the error estimate. We consider the example of \kcorr\ from observed \rp\ band to rest-frame $B$ band in Figure~\ref{fig:kobsv}.

The left panel plots the comparison of statistical \kcorr\ errors calculated using the revised template of N02 and the new template. The result is similar to that of Figure~\ref{fig:krest}. Since the spline knots are placed at different locations for the two methods of color-correction, the similarity between the results gives us confidence that the method of constructing spectral templates presented in Section~\ref{sec:template} is largely insensitive of the way the spectra are color-corrected.

Color-correcting using observed colors has the advantage of being independent of an assumed stretch-color relation, but the \kcorr\ errors depend on the availability of observed colors. The effects of the availability of observed colors are shown in the right panel of Figure~\ref{fig:kobsv}. We consider three cases where the new template is color-corrected by both observed $\gp-\rp$ and $\rp-\ip$ colors, by observed $\gp-\rp$ color and by observed $\rp-\ip$ color. They illustrate the point that in the treatment of \kcorr, the errors should also reflect the availability of observed colors. For example, the \kcorr\ error is increased considerably for a SN\,Ia missing $\gp-\rp$ observed color at $z=0.36$ (dashed curve of the right panel of Figure~\ref{fig:kobsv}). A similar effect is also observed for a SN\,Ia at $z=0.54$ missing $\rp-\ip$ (dotted curve of the right panel of Figure~\ref{fig:kobsv}). 

\begin{figure}
\plotone{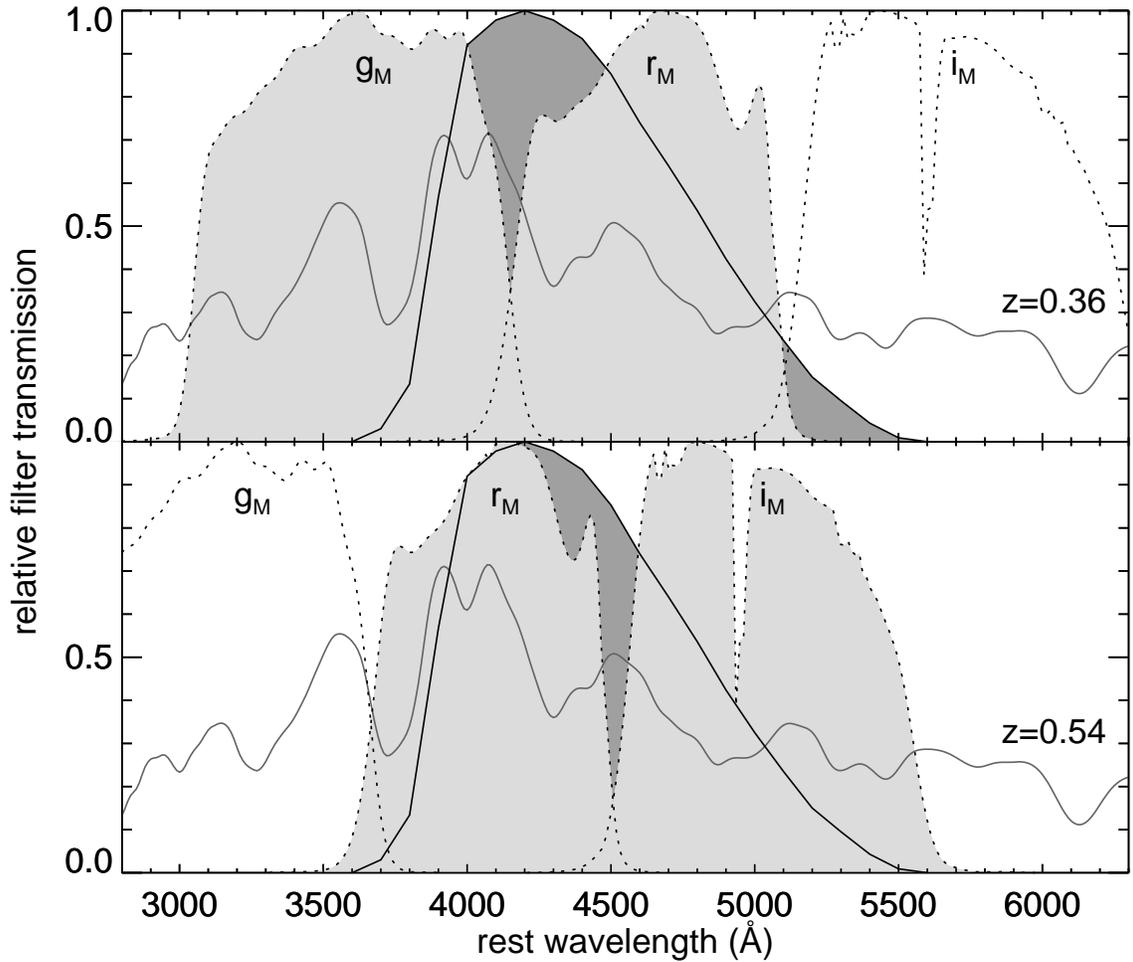}
\caption{The location of the observed filters at redshifts $z=0.36$ (top panel) and $z=0.54$ (bottom panel). We consider the same example of \kcorr\ as in Figure~\ref{fig:kobsv}. The solid curves represent the rest-frame $B$ band, while the dotted curves represent the observed \gp\rp\ip\ bands. At redshifts where the observed and rest-frame filter bands are misaligned, the observed color which straddle the rest-frame filter band (highlighted for each plot) is the most important.
  \label{fig:obsvcol}}
\end{figure}

At redshifts where the observed and rest-frame filter bands are misaligned, the observed color which straddles the rest-frame filter band is the most important. This is illustrated in Figure~\ref{fig:obsvcol}. For a SN\,Ia that is at a misaligned redshift, and missing this observed color information, the associated \kcorr\ error arises not only from inhomogeneity of SNe\,Ia spectral features, but also from the missing color information.

\subsection{Determining the weights from the minimization of errors}
\label{sec:error:errormin}

In Section~\ref{sec:template:weight}, we selected two constants for the weighting scheme, constants for $w_{type}$ and $w_{coverage}$, in order to ensure that the spectral template produced is not dominated by an extreme of the SN\,Ia population. We used a constant of $0.2$ for $w_{type}$ to simulate the fraction of peculiar supernovae in the SN\,Ia population. We used a constant of $0.2$ for $w_{coverage}$ to weight down the edges of the spectra. Alternatively, these constants can be quantitatively determined by minimizing the systematic \kcorr\ errors of a set of spectra.

We set the constants for $w_{type}$ and $w_{coverage}$ as parameters which are free to vary between $0$ and $1$. We took all of the spectra which have enough wavelength coverage to allow for accurate mangling to determine the systematic errors in redshift as discussed in Section~\ref{sec:error:rest}. The parameters are quantitatively determined by minimizing these errors using non-linear least-squares fitting \citep{Mor80}. At each iteration, a spectral template is produced from the entire spectral library with the specified parameters, and the averages of the \kcorr\ differences between this template and the library spectra are taken as the systematic errors in redshift to be minimized.

By minimizing the systematic errors, the optimal constants for $w_{type}$ and $w_{coverage}$ were determined to be $0.244$ and $0.498$, respectively. The constant for $w_{type}$ roughly reflects the fraction of peculiar spectra in the sample. The larger constant for $w_{coverage}$ reflects the fact that the edges of these library spectra are used to determine the errors at redshifts where the filters are misaligned.

\begin{figure}
\plotone{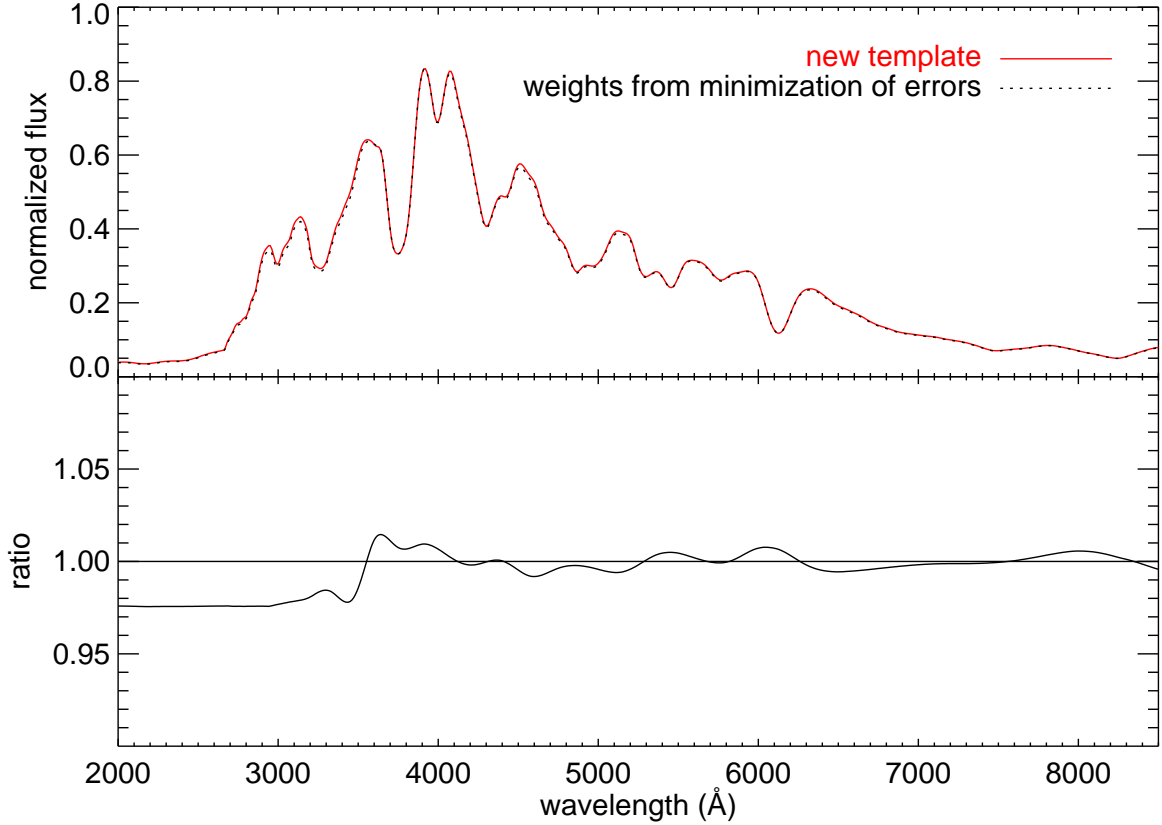}
\caption{Comparison between the spectral template built using weights determined by minimizing the \kcorr\ errors and preselected weights. The templates at maximum light are shown. The template building process is largely insensitive to the way these constants are determined. The largest \kcorr\ (to $B$) difference between the two templates is $0.004$.
  \label{fig:kerrormin}}
\end{figure}

The spectral template produced by using the quantitatively determined parameters shows very little difference from the spectral template presented in Section~\ref{sec:template}. The templates at maximum light are compared in Figure~\ref{fig:kerrormin}. The largest \kcorr\ (to $B$) difference between the two templates is $0.004$. Even when the two constants are set as their maximum value of unity, the maximum \kcorr\ difference is still less than $0.01$. The spectral template building process is therefore largely insensitive to the way these constants are determined.

The statistical error estimates remain on the order of $0.01$ at the redshifts where the filters are aligned and $0.04$ at the redshifts where the filters are misaligned. The minimization of the systematic error represents an improvement of $\sim 0.0005$ at the misaligned redshifts and has no effects on the shape of the systematic errors in redshift. The insensitivity to these parameters is mostly due to the fact that our spectral library is large and encompasses SNe\,Ia which span a wide range of properties.

Despite the fact that the template building process is largely unaffected by the choice of these parameters, the weights determined from the minimization of errors depend on the characteristics of the sample used for the error estimates. To use the method presented in this section, one should ensure that the sample of spectra has good spectral and epoch coverage and is reasonably representative in properties such as the stretch and the fraction of peculiar spectra. Using an unrepresentative sample of spectra has the potential to skew the template toward a particular extreme of the SN\,Ia population.


\section{Impact of \kcorr\ errors on cosmology}
\label{sec:cosmology}

In the preceding sections we have concentrated on determining the \kcorr\ errors for individual observations. Since one generally observes a supernova at multiple epochs, it is natural to ask what the final effect of these errors will be on the derived light-curve parameters, such as the peak magnitude in $B$.

This question is currently difficult to answer. The critical issue is how correlated the \kcorr\ error is at different epochs. If the errors were completely uncorrelated, as is often assumed in the literature, then the final error would be approximately reduced by $1/\sqrt{N}$, where $N$ is the number of observations (this is not quite true because different observations tend to have different weights in a light-curve fit). Alternatively, if the \kcorr\ errors were perfectly correlated, then there would be no reduction in the final error from multiple observations.

In reality, we are somewhere between these extremes. It is easy to understand why observations at different epochs would be correlated. For example, imagine a supernova at a redshift where a particular feature is barely outside a given observed filter for the fiducial SED. If this supernova has unusually high expansion velocities, then the feature might be shifted just inside the observed filter, which would affect the \kcorr. Since a supernova with a high expansion velocity relative to the fiducial template is likely to also have a high relative expansion velocity at a later date, the same effect will occur when the supernova is observed several days later. Thus, the errors in the \kcorr\ will be correlated between different observations.

What is needed to determine the level of correlation is SNe\,Ia with spectra obtained at many epochs. One can then calculate the \kcorr\ errors for the same supernova at different epochs and form the covariance matrix. Unfortunately, at the current time there are not nearly enough SNe\,Ia with multiple published spectra. New data samples such as those which should be provided by the Carnegie Supernova Project \citep{2006PASP..118....2H}, Katzman Automatic Imaging Telescope \citep{2000AIPC..522..103L}, the Supernova Factory \citep{2004NewAR..48..637W} and the European Supernova Collaboration as part of a European Research Training Network \citep[e.g.,][]{Pas07} will help address this issue.

Nonetheless, we can try to develop a feel for the importance of this effect with the current data, although this will necessarily be highly speculative. We took all of the SNe\,Ia with multiple spectra which have enough wavelength coverage to allow for accurate mangling, calculated their \kcorr\ errors as discussed in Section~\ref{sec:error:rest}, and formed the covariance matrix by taking the appropriate products. This covariance matrix has 275 unique off-diagonal elements (from 25 epoch bins defined in Table~\ref{tbl:epoch}). The current data set provides information for only 10\% of these, many of which have entries from only one supernova.

\begin{figure}
\plotone{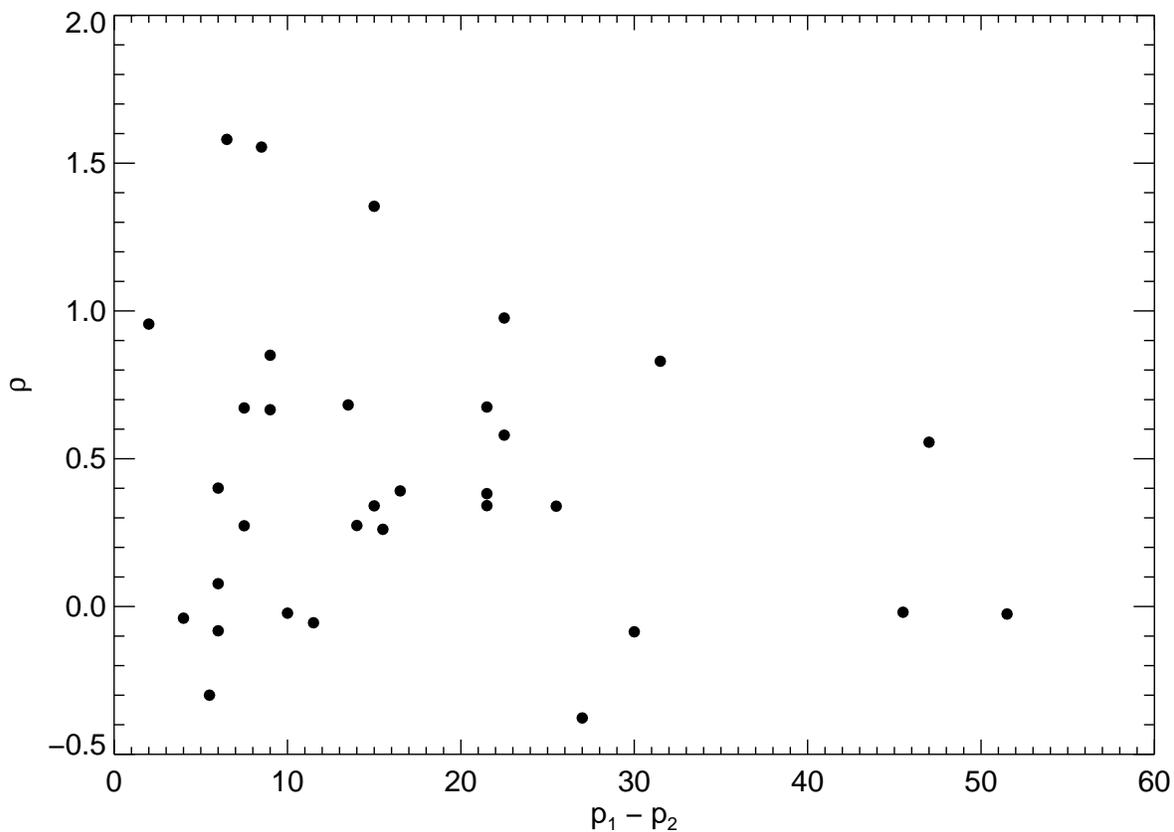}
\caption{Estimated correlation coefficients for the $\sim 10\%$ of the off-diagonal elements of the \kcorr\ covariance matrix that we can currently measure plotted against the difference between the two epochs. Here the $z=0.8$ $\ip \mapsto B$ is shown. At this redshift the filters line up fairly well. Note that some of the correlation coefficients are greater than one, which is unphysical, and results from the fact that many of these bins have only one SN\,Ia in them.
\label{fig:corr_coeffs}}
\end{figure}

Clearly, with only 10\% of the elements specified, we cannot estimate the final error without making some additional coefficients. The distribution of correlation coefficients for the \kcorr\ from \ip\ to $B$ at $z=0.6$ is shown in Figure~\ref{fig:corr_coeffs}, plotted against epoch difference between the two bins. Note that some of the correlation coefficients are larger than 1, which is non-physical and is related to the fact that we only have one supernova in that bin from which to estimate the covariances. There are no clear trends visible, which is not surprising given the paucity of data.

Therefore, to proceed, and for lack of any better model, we measure the mean correlation coefficient ($\sim 0.5$) and use that to calculate all of the off-diagonal elements. We carry this operation out at three redshifts ($z = 0.6, 0.7, 0.8$), in all cases calculating the error in the \kcorr\ from \ip\ to $B$. The alignment between $B$ and \ip\ is best at $z=0.8$. In order to make this estimate, we have to have some idea how \ip\ band observations are typically distributed at these redshifts. For this we use three actual SNLS SNe\,Ia, 03D4dy ($z=0.604$), 04D1si ($z=0.701$), and 04D1ks ($z=0.798$). At higher redshifts, we actually tend to have slightly more measurements because of time dilation.

We then calculate the expected error on the peak $B$ magnitude in three different ways: assuming that the errors are completely uncorrelated, using only the covariance matrix elements we have actually measured (i.e., with 10\% of them non-zero), and using the above model for the unknown coefficients. The results are summarized in Table~\ref{tbl:final_error}. Roughly speaking, it appears that the errors on the final magnitude are a factor of 2 or so larger than one would expect in the absence of correlations. We must emphasize that these numbers are tentative, but they should be a considerable improvement over the assumption that the \kcorr\ errors at different epochs are uncorrelated.

\begin{figure}
\plotone{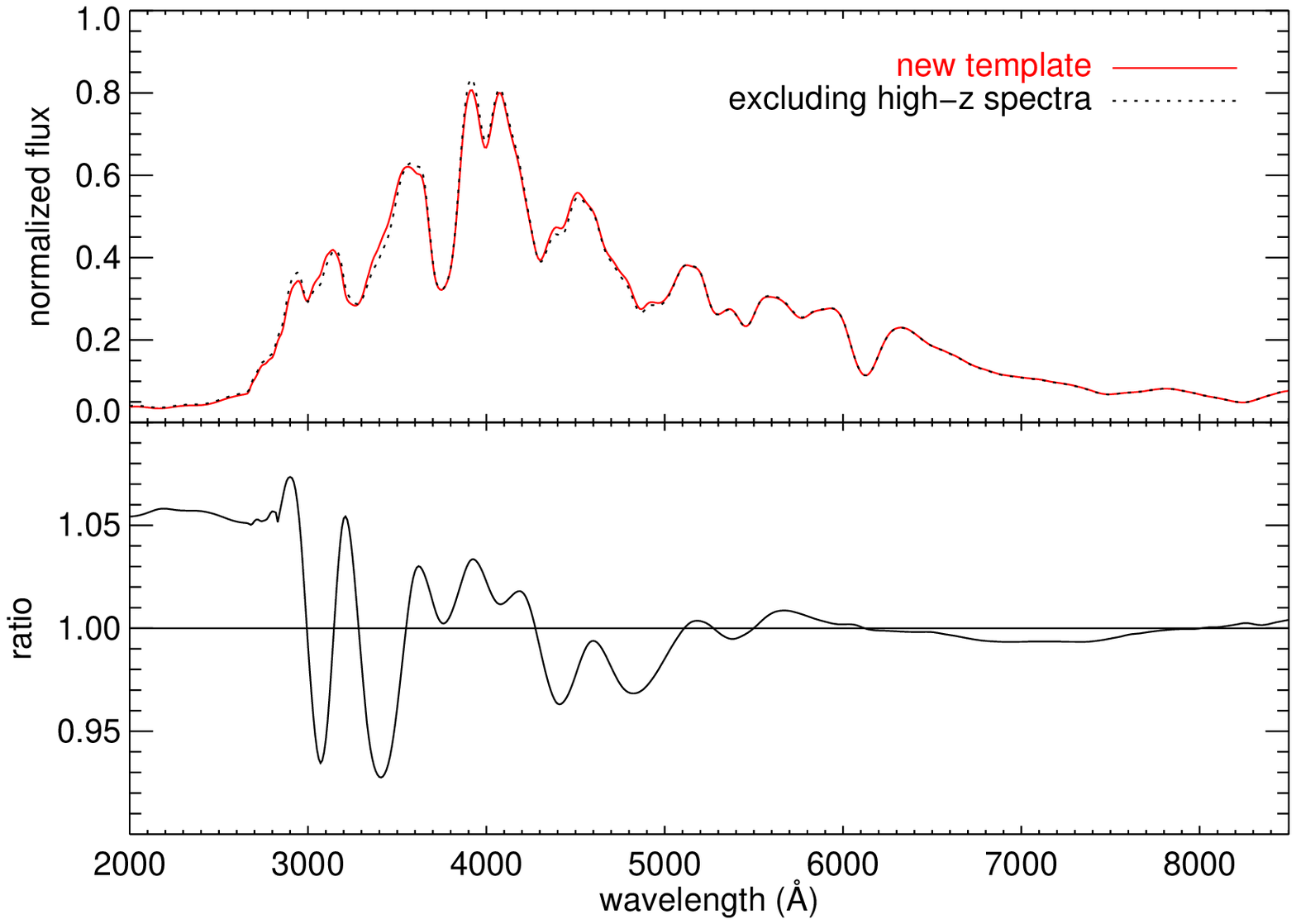}
\caption{Comparison between the spectral template built with the full sample of library spectra (solid curve) and the spectral template built with only low-redshift spectra (dotted curve). The templates at maximum light are shown. The difference in \kcorrs\ between the two spectral templates is small and mostly reflects the difference in the wavelength coverage of the two samples rather than true evolution.
  \label{fig:hiloz}}
\end{figure}

Differences between local and high-redshift SNe\,Ia are also of interest in cosmology. The detailed comparison of spectra of nearby and distant SNe\,Ia is the subject of many papers \citep[e.g.,][]{2006AJ....131.1648B,Bro07}. In the previous sections, we have assumed that the effect of the evolution in redshift is negligible compared to the intrinsic variations in spectral features between SNe\,Ia at the same redshift. Here, we perform a basic test on the impact of including both low-redshift ($z<0.2$) and high-redshift samples in the template building process. We build the spectral template using only low-redshift spectra to compare with our new spectral template built using the full sample of library spectra (Figure~\ref{fig:hiloz}). The difference in \kcorrs\ between the two spectral templates is small (typically 0.01 to rest-frame $B$ band) and mostly reflects the difference in the wavelength coverage of the two samples rather than true evolution.


\section{Template spectroscopic sequence}
\label{sec:sequence}

It is possible to establish a one-to-one relation between light-curve shape and spectral feature strengths \citep{1995ApJ...455L.147N,2005ApJ...623.1011B,2006MNRAS.370..299H} for certain features around the epoch of maximum $B$ band light. Quantifying the relative feature strengths of a large sample of SN\,Ia spectra (Section~\ref{sec:template:measure}) allows us to explore the possibility of constructing a template spectroscopic sequence as a function of the light-curve shape.

As a first test, we consider the well known spectroscopic sequence of Si II $\lambda5972$ to Si II $\lambda6355$ ratio around $B$ band maximum light first explored by \citet{1995ApJ...455L.147N}. We select 41 library spectra near maximum light (one day before to one day past maximum light) with adequate $V$ band coverage. The spectra are of 28 SNe\,Ia with available light curves and stretch factors.

We perform a principal component analysis (PCA) on the feature strength measurements to characterize the spectral features in this region. PCA reduces multidimensional data to principal components with fewer dimensions by identifying patterns in the data. PCA has been employed for analyzing astronomical spectra in a wide variety of applications \citep[e.g.,][]{1999MNRAS.303..284R,2001MNRAS.323.1035K,2003ApJ...599L..33M,2006MNRAS.370..828F}. Here, PCA is applied to our color-independent feature strength measurements (Section~\ref{sec:template:measure}) of the selected spectra. 

We consider the spectral region in the $V$ band (4600\AA\ to 6600\AA). Each spectrum is characterized by nine feature strength measurements. Using the feature strength measurements instead of the observed spectra themselves minimizes the effects of noise in the analysis. The first principal component is taken along the direction of maximum variance in the nine-dimensional space and constitutes $\sim60\%$ of the total variance. We assume here that the first principal component describes the pattern instigated by the variation in light-curve shapes (e.g., stretch). We establish a one-to-one relation between the pattern described by the first principal component and the stretch factors of the SNe\,Ia and derive a spectroscopic sequence.

\begin{figure}
\plotone{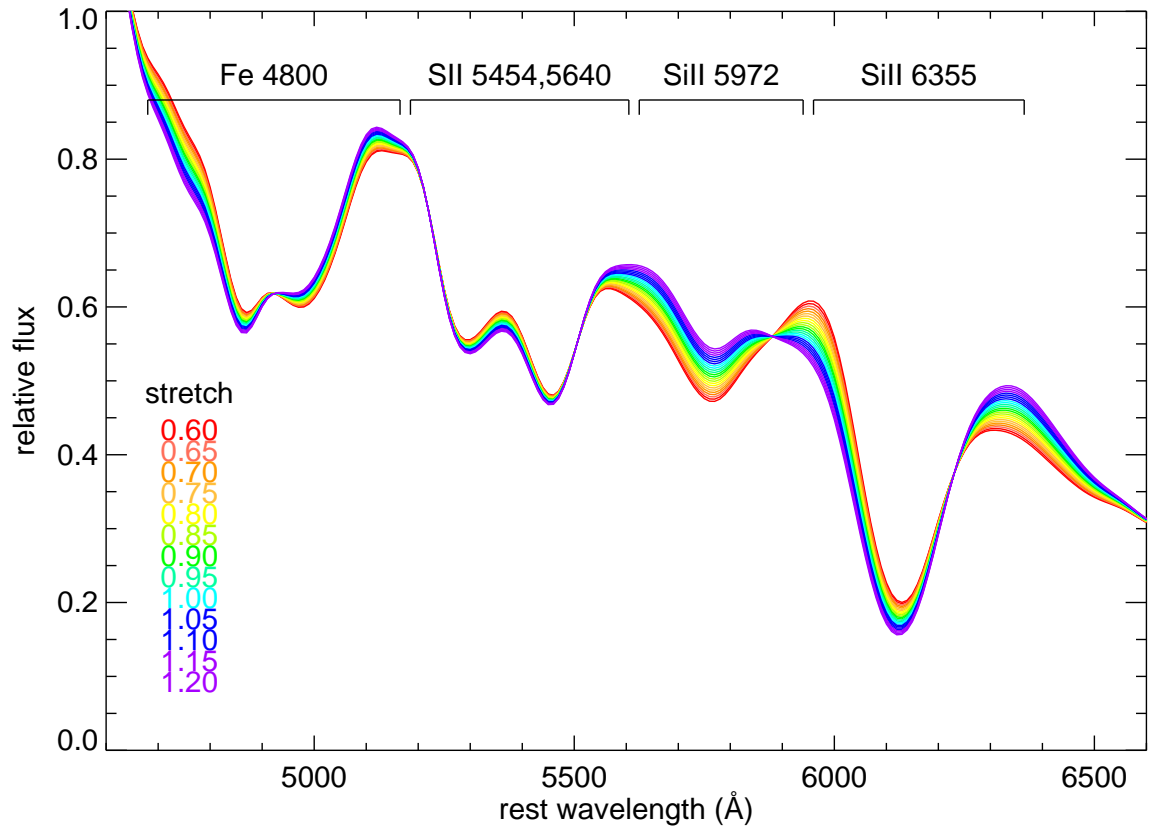}
\caption{The template spectroscopic sequence in the $V$ band at maximum light. The template is derived from 41 library spectra of 28 SNe\,Ia with a wide range of stretch factors. The pattern is prescribed by the principal component analysis of the narrow-band color measurements of the library spectra.
  \label{fig:seq}}
\end{figure}

The resulting template spectroscopic sequence as a function of wavelength and stretch factor is plotted in Figure~\ref{fig:seq}. The color dependence of SN\,Ia spectra on their stretch factors is removed by color-correcting all the spectra to the same broad-band colors. Figure~\ref{fig:seq} then exhibits the spectral variations due to the varying stretch factor independent of colors. The feature strength ratio of Si II $\lambda5972$ to Si II $\lambda6355$ decreases and the equivalent width of the Fe trough near 4800\AA\ trough stays roughly constant as the stretch factor increases (as the declining rate decreases). This description of the spectroscopic sequence agrees with the findings of \citet{2006MNRAS.370..299H}. It represents an approximately $0.005$ mag reduction in error for a \kcorr\ from \ip\ to $V$ at a misaligned redshift of $z=0.5$ for a SN\,Ia with an extreme stretch value. The improvement is small compared to the $0.04$ mag statistical error caused by feature differences which cannot be predicted by light-curve shapes.

A spectroscopic sequence as a function of stretch factor is possible only for a limited range of epochs and spectral regions. SNe\,Ia with the same stretch factor and intrinsic colors in general do not have identical spectral features at all wavelengths and at all epochs. Some SNe\,Ia with extreme stretch values sometimes exhibit ``normal'' spectral features, while other SNe\,Ia with moderate stretch values sometimes are identified as spectroscopically peculiar  \citep{2004AJ....128..387G}. The spectral features of SNe\,Ia clearly depend on more parameters than the stretch factor alone. This again emphasizes the need to characterize the \kcorr\ errors with as large a sample of supernova spectra as possible.


\section{Conclusions}
\label{sec:conclusions}

In the era of large SNe\,Ia magnitude-redshift surveys, the determination of \kcorrs\ and their errors becomes especially important. While \kcorrs\ are largely driven by the colors of a SN\,Ia, it is shown here that spectral features also play a significant role. The statistical errors caused by the diversity in spectral features are measured to be $0.04$ mag for a single observation at a redshift where large extrapolation is required.

We have presented in this paper improvements in the \kcorr\ calculations and in the error estimates. The method for color-correction simultaneously allows for the adjustment of multiple colors and a consistent continuum for the comparison of spectral features between spectra. The method of deriving a representative spectral template allows the inclusion of a large sample of observed spectra and reduces the systematic errors of \kcorrs. The redshift-dependent nature of \kcorr\ errors derived here shows that SNe\,Ia at different redshifts should be weighted differently in the determination of cosmological parameters. 

\acknowledgments 
This paper is based in part on data gathered with the 6.5 meter Magellan Telescopes located at Las Campanas Observatory, Chile. It is also based on observations obtained at the Gemini Observatory and the W.~M.~Keck Observatory on Mauna Kea. The authors would like to recognize the very significant cultural role and reverence that the summit of Mauna Kea has within the indigenous community of Hawai'i. We are grateful for our opportunity to conduct observations from this mountain. Canadian authors acknowledge support from the Natural Sciences and Engineering Research Council of Canada and the Canadian Institute for Advanced Research. This research has made use of the NASA/IPAC Extragalactic Database which is operated by the Jet Propulsion Laboratory, California Institute of Technology, under contract with the National
Aeronautics and Space Administration. This work is also based in part on Carnegie Supernova Project observations supported by the National Science Foundation under Grant No. 0306969.




\begin{deluxetable}{ccccc}
\tablecaption{The assignment of epoch bins \label{tbl:epoch}}
\tablewidth{0pt}
\tablehead{\colhead{Lower epoch} & \colhead{Upper epoch} & \colhead{Bin size} & \colhead{Effective epoch} & \colhead{Number of spectra}}
\startdata
-15 & -10 & 6 & -11.6 & 18 \\
-9 & -8 & 2 & -8.4 & 24 \\
-7 & -6 & 2 & -6.5 & 29 \\
-5 & -4 & 2 & -4.5 & 24 \\
-3 & -2 & 2 & -2.5 & 23 \\
-1 & 0 & 2 & -0.4 & 43 \\
1 & 2 & 2 & 1.5 & 35 \\
3 & 4 & 2 & 3.4 & 25 \\
5 & 6 & 2 & 5.5 & 22 \\
7 & 8 & 2 & 7.6 & 30 \\
9 & 10 & 2 & 9.5 & 21 \\
11 & 12 & 2 & 11.5 & 24 \\
13 & 15 & 3 & 14.1 & 27 \\
16 & 18 & 3 & 16.6 & 26 \\
19 & 21 & 3 & 19.7 & 22 \\
22 & 24 & 3 & 22.9 & 25 \\
25 & 28 & 4 & 26.8 & 22 \\
29 & 32 & 4 & 30.8 & 26 \\
33 & 37 & 5 & 34.9 & 23 \\
38 & 41 & 4 & 39.5 & 27 \\
42 & 49 & 8 & 44.9 & 24 \\
50 & 56 & 7 & 53.0 & 24 \\
57 & 67 & 11 & 61.0 & 21 \\
69 & 79 & 11 & 73.6 & 10 \\
81 & 91 & 11 & 84.9 & 13 \\
\enddata
\end{deluxetable}


\begin{deluxetable}{lccc}
\tablecaption{\kcorr\ errors in peak magnitudes}
\tablehead{
  \colhead{ } & \colhead{ z = 0.6 } & \colhead{ z = 0.7 } &
  \colhead{ z = 0.8 }
}
\startdata
  Uncorrelated         & 0.011 & 0.011 & 0.004 \\
  Partially correlated & 0.013 & 0.011 & 0.006 \\
  Correlations model   & 0.023 & 0.018 & 0.010 \\
\enddata
\tablecomments{The approximate effects of including correlations between \kcorrs\ on different epochs on the peak magnitude in $B$. The uncorrelated numbers reflect the assumption that the \kcorr\ errors are completely uncorrelated, which is unlikely to be true as explained in the text. The partially correlated model only includes those correlations we can estimate from current data ($\sim 10\%$), and the correlations model reflects the effect of very roughly estimating the unknown correlations using a simple model (setting them all equal to the mean value we can estimate).
 \label{tbl:final_error}
}
\end{deluxetable}

\end{document}